\documentclass[12pt, pre, preprint]{revtex4-1}
\usepackage{graphicx, amsmath, amssymb, amsfonts}
\begin{document}
\title{Anomalous local coordination, density fluctuations, and void statistics in disordered hyperuniform many-particle ground states}
\author{Chase E. Zachary}
\email{czachary@princeton.edu}
\affiliation{Department of Chemistry, Princeton University, Princeton, NJ 08544}
\author{Salvatore Torquato}
\email{torquato@princeton.edu}
\affiliation{Department of Chemistry, Department of Physics, 
Program in Applied and Computational Mathematics, 
Princeton Institute for the Science and Technology of Materials, and Princeton Center for Theoretical Science, Princeton University, Princeton, NJ 08544}
\date{\today}

\begin{abstract}
We provide numerical constructions of one-dimensional hyperuniform 
many-particle distributions that exhibit unusual clustering and 
asymptotic local number density fluctuations growing more 
slowly than the volume of an observation window but faster than the surface area.  
Hyperuniformity, defined
by vanishing infinite-wavelength local density fluctuations, 
provides a quantitative metric of global order within a 
many-particle configuration and signals the onset of an ``inverted'' critical point in which the 
direct correlation function becomes long-ranged.  By targeting 
a specified form of the structure factor at small wavenumbers 
using collective density variables, we are able to 
tailor the form of asymptotic local density fluctuations
while simultaneously measuring the effect of imposing weak 
and strong constraints on the available degrees of freedom within the system.  
This procedure is equivalent to finding the (possibly disordered) 
classical ground state of an interacting many-particle system with up to four-body interactions.
Even in one dimension, the long-range effective interactions induce 
clustering and nontrivial phase transitions in the resulting ground-state configurations.
We provide an analytical connection between the fraction of 
contrained degrees of freedom within the system and the disorder-order 
phase transition for a class of target structure factors by examining the realizability of 
the constrained contribution to the pair correlation function.
Our results explicitly demonstrate that disordered hyperuniform 
many-particle ground states, and therefore also point distributions, 
with substantial clustering can be constructed.
We directly relate the local coordination structure of 
our point patterns to the distribution of the void space external to the 
particles, and we provide a scaling argument for the 
configurational entropy of the systems when only a small fraction of the 
degrees of freedom are constrained.  
By emphasizing the intimate connection between 
geometrical constraints on the particle distribution and structural regularity, 
our work has direct implications for higher-dimensional systems, 
including an understanding of the appearance of hyperuniformity and 
quasi-long-range pair correlations in maximally random strictly jammed packings of hard spheres.
\end{abstract}

\maketitle

\section{Introduction}

The relationship between the local structure of a many-particle system 
and interparticle correlations is fundamental to condensed-matter theory.  
This intimate connection 
provides a useful image of the regularity \cite{FN1a} of all phases of matter, 
allowing researchers to track the local structure over increasing length scales 
approaching the global system.  In practice, one measures pair correlations
between distinct points in the form of the structure factor $S(k)$, 
which is proportional to the scattering intensity from 
x-ray or small-angle neutron scattering \cite{Ch87}.  
It is intuitive from such measurements that a hierarchy of structural order
can be established, ranging from crystalline structures such as Bravais 
lattices \cite{FNBV} to highly disordered systems, 
the prototypical example of which is the ideal 
gas \cite{ToSt03, ToTrDe00, ZaTo09, StNeRo83}.  
Unfortunately, quantitative descriptors consistent with this 
stratification of order are difficult to identify, and this area of 
research is currently open.  One recently introduced order 
metric \cite{ToSt10RMP} involves the notion of hyperuniformity 
of point patterns, whereby infinite-wavelength local density fluctuations
vanish \cite{ToSt03, ZaTo09}.  This order metric explicitly indicates 
the degree to which density fluctuations are suppressed on large length scales.  

The local structure of a hyperuniform many-particle configuration 
(i.e., on the order of a few nearest-neighbor distances between particles) 
is by definition indicative of the global arrangement of particles \cite{ToSt03}.
Also known as superhomogeneity \cite{PiGaLa02}, this phenomenon 
is fundamental to the description of all Bravais lattices, lattices with 
a multiparticle basis, quasicrystals, 
and certain disordered systems possessing pair correlation functions 
decaying to unity exponentially fast \cite{ZaTo09}.  
We emphasize that while hyperuniformity in periodic configurations 
is a trivial consequence of their intrinsic long-range order, the fact 
that disordered many-particle systems can also display this property 
is nonintuitive.  
This behavior is especially surprising since the appearance of 
hyperuniformity marks the onset of an ``inverted'' critical point in 
which the structure factor 
vanishes in the limit of small wavenumbers while the direct correlation function, 
defined through the Ornstein-Zernike formalism, becomes long-ranged \cite{ToSt03}.  

Hyperuniform systems have played a fundamental role in our 
understanding and design of materials, including those with large, 
complete photonic band gaps \cite{FlToSt09}, ``stealth'' materials invisible 
to certain frequencies of radiation \cite{BaStTo08}, and prototypical glassy 
structures consisting of maximally random 
strictly jammed (MRJ) monodisperse hard spheres \cite{DoStTo05, ZaJiTo10}.  
Other examples of disordered hyperuniform systems include noninteracting 
spin-polarized fermions \cite{ToScZa08, ScZaTo09}, the ground state of 
liquid helium \cite{ReCh67}, 
the density fluctuations of the early Universe \cite{Pe98}, one-component 
plasmas \cite{ToSt03}, and so-called $g_2$-invariant processes \cite{ToSt03}, 
in which the form of the pair correlation function is held fixed over a 
certain density interval.   
Note for \emph{equilibrium} many-particle configurations at positive 
temperature, hyperuniformity implies that the isothermal 
compressibility vanishes; this relationship does not hold, however, 
for nonequilibrium systems.

Hyperuniform particle distributions possess structure factors with a 
small-wavenumber scaling $S(k) \sim k^{\alpha}$ for $\alpha > 0$, 
including the special case $\alpha = +\infty$ for periodic crystals. 
This behavior implies that the variance $\sigma^2_N(R)$ in the 
number of particles within a local observation window 
(here a $d$-dimensional sphere of radius $R$) increases 
asymptotically as \cite{ZaTo09}
\begin{equation}\label{NVscaling}
\sigma^2_N(R) \sim \begin{cases}
R^{d-1}\ln R, & \alpha = 1\\
R^{d-\alpha}, & \alpha < 1\\
R^{d-1}, & \alpha > 1
\end{cases}\qquad (R\rightarrow +\infty).
\end{equation}
However, all known hyperuniform configurations to date 
have a scaling parameter $\alpha \geq 1$ \cite{UcToSt06, GaJoTo08}, 
meaning that the second asymptotic regime of the number variance 
in \eqref{NVscaling} has never been observed in either theoretical or 
experimental studies.  
Indeed, the aforementioned MRJ packings, which are \emph{maximally} 
disordered among all jammed sphere packings with diverging elastic moduli, 
possess a small-wavenumber scaling $\alpha = 1$, and this observation
has provoked the question of whether this value corresponds to a \emph{minimal} 
scaling among all hyperuniform point patterns.  Zachary, Jiao, and Torquato have
 provided strong arguments that 
this claim is indeed true for strictly jammed hard-particle packings \cite{ZaJiTo10},
but it is unclear whether general point patterns must also possess 
exponents $\alpha \geq 1$. 
Here we provide for the first time constructions of ``anomalous'' 
disordered hyperuniform many-particle ground states for which $\alpha < 1$, 
demonstrating the diversity of possible structures within of this class of systems.  

Our approach involves placing explicit constraints on the so-called collective 
coordinates associated with a point distribution, which are defined 
by a Fourier transform of the local density variable 
(discussed in Section II below) \cite{FaPeStSt91, UcStTo04,UcToSt06}. 
Controls on collective coordinates have been previously used 
in the development of novel stealth materials \cite{BaStTo08} and 
in the identification of unusual disordered classical ground states 
for certain classes of pair potentials \cite{BaStTo09}.
This problem can be viewed as the determination of the ground state 
of a many-particle system with up to four-body interactions \cite{UcToSt06}; 
duality relations that relate the energy per particle of a many-body 
potential in real space 
to the corresponding energy of the dual (Fourier-transformed) potential 
can be used to examine analytically the ground state structures and 
energies \cite{ToSt08}. 
Importantly, since collective coordinates directly probe the configuration 
space associated with the two-particle information of the structure factor, 
they are ideally suited to the construction of hyperuniform point patterns.    
Formally, we numerically construct a configuration of particles 
whose spatial distribution is consistent with a targeted form of the structure 
factor at small wavenumbers.  By constraining a certain number of degrees of 
freedom in the system, we ``fix'' the positions of a known fraction of the total 
number of particles
based on the locations of the remaining particles and the implicit constraints 
imposed by the targeted form of $S(k)$.  
By varying the fraction of constrained degrees of freedom 
within the system, we are able to explore directly the relationship between 
hyperuniformity and internal structural constraints of a many-particle configuration, 
allowing us to interpolate between the ``disordered'' and ``ordered'' regimes of 
hyperuniformity.  

\begin{figure}[!t]
\centering
\includegraphics[width=0.45\textwidth]{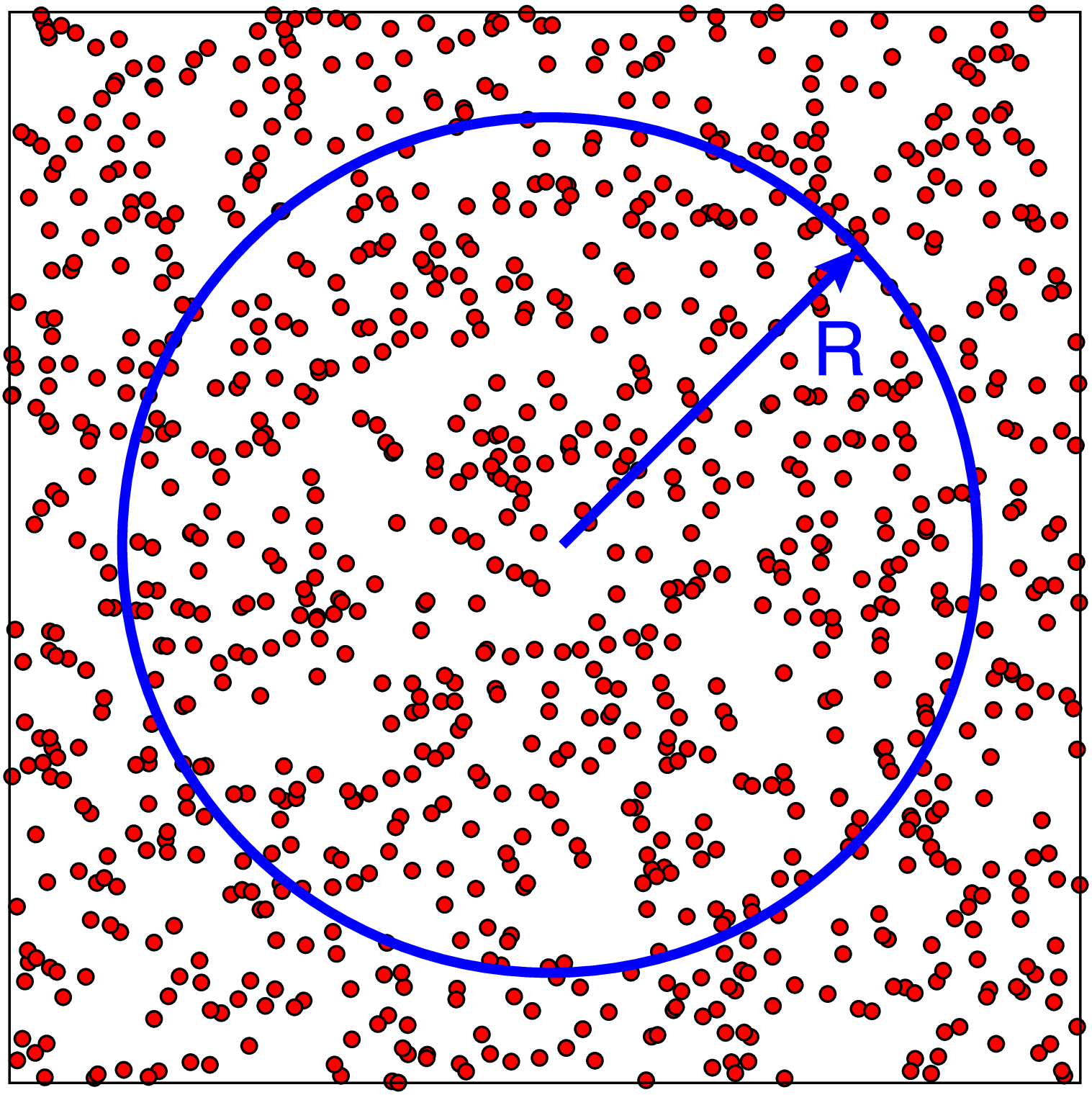}\hspace{0.05\textwidth}
\includegraphics[width=0.45\textwidth]{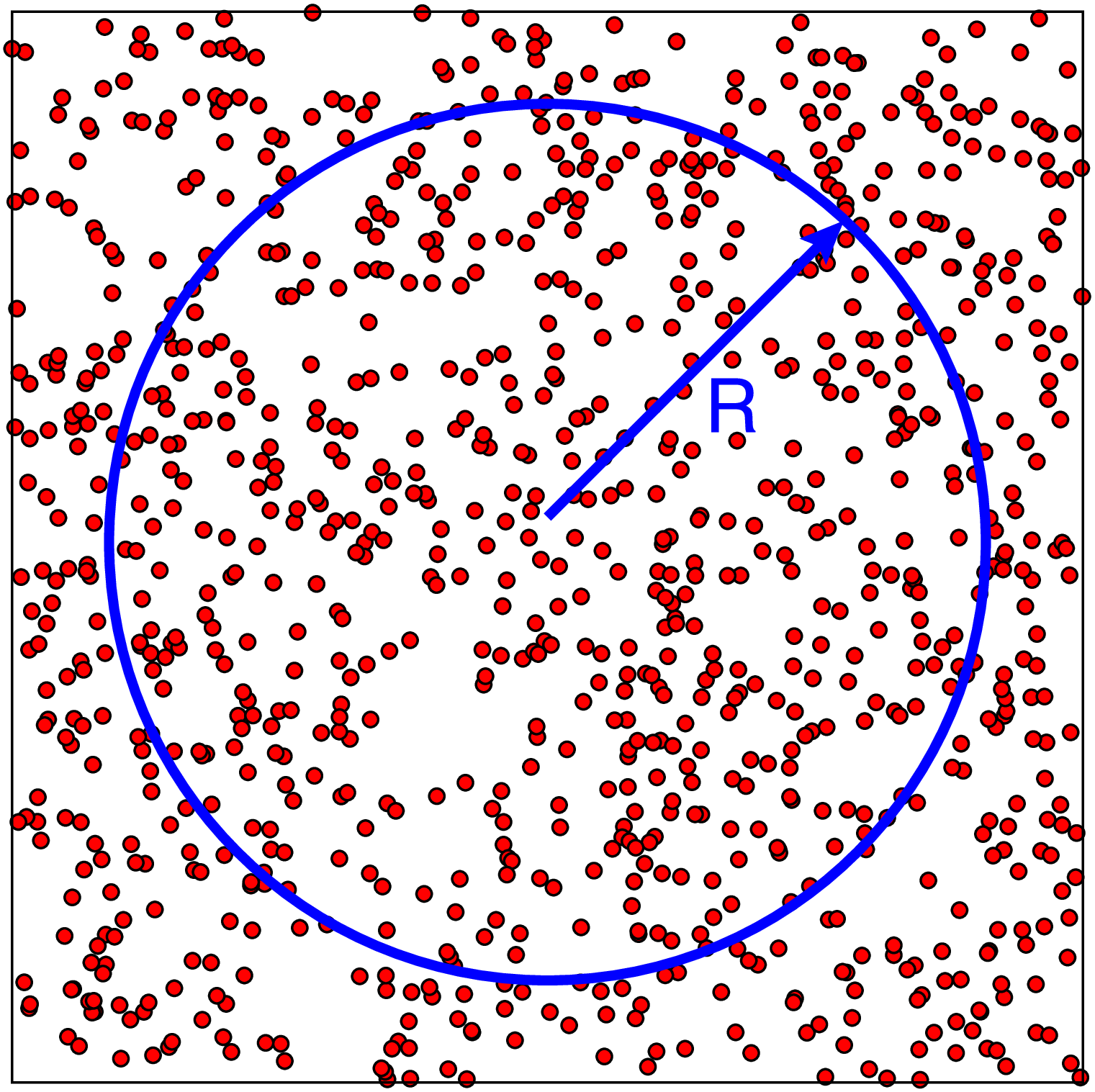}
\caption{(Color online)  Numerically-generated configurations 
of particles in two dimensions with a circular local observation window of radius $R$.  
Both configurations exhibit strong local clustering of points and 
possess a highly irregular local structure; however, 
the configuration on the left is hyperuniform while the one on the right is not.  
The hyperuniform point pattern was generated with the same methodology 
outlined in Section III of the text.  }\label{PPfig}
\end{figure}
In order to elucidate the connection between the local coordination structure and 
pair correlations for our anomalous hyperuniform ground states, we have 
investigated the distribution of the available \emph{void space} external to the particles.  
Prior work on MRJ packings 
of binary hard disks has shown that the appearance of hyperuniformity in a 
many-particle system is related to the underlying distribution of the local voids 
between particles \cite{ZaJiTo10}; 
in this sense, the void space is more fundamental to the local structure than 
the particles themselves.  Strong arguments have also been put forth to support 
the claim that exponential values $\alpha$ less than unity in the small-wavenumber 
region of the structure factor indicate the presence of larger interparticle voids
with higher frequency, thereby deregularizing the microstructure while maintaining 
hyperuniformity \cite{ZaJiTo10}.  This behavior is notable since it is not obvious that 
hyperuniformity can be consistent with a highly clustered microstructure; 
see Fig. \ref{PPfig}.   
Here we provide further evidence to link rigorously the void space and the local 
coordination structure of a point pattern, and we highlight the differences in the void space 
distribution for ``regular'' and ``anomalous''  hyperuniform systems.  
Since we can directly control the fraction of constrained degrees of freedom via 
collective coordinates, our results have implications for understanding how the void space 
distribution is affected by increased constraints on the many-particle configuration. 
 Indeed, our work directly supports the fundamental role of the void space in the 
 microstructure and reinforces the relationship between constraints on the local structure
and the aforementioned observed minimal scaling $\alpha = 1$ found in $S(k)$ for 
MRJ hard sphere packings.  

Our major results are summarized as follows:
\begin{itemize}
\item[(i)]  Disordered hyperuniform many-particle ground states can, counterintuitively, 
exhibit a substantial degree of clustering in the absence of a large number of constraints 
on the particle distribution (Sections IV and V).
\item[(ii)]  The order-disorder phase transition that occurs upon increasing the fraction of 
constrained degrees of freedom is related to the \emph{realizability} of the constrained 
contribution to the pair correlation function $g_2(r)$ (defined below)
(Section IV).
\item[(iii)]  Hyperuniform particle distributions with anomalous asymptotic local density 
fluctuations (i.e., slower than the volume but faster than the surface area of an observation 
window) can be constructed, and these fluctuations are intimately
related to the distribution of \emph{void sizes} external to the particles (Section V).
\item[(iv)]  With few constrained degrees of freedom (e.g., a perturbation from an ideal gas), 
the entropy (configurational degeneracy) decreases linearly with the number of constraints 
imposed on the particle distribution (Section V).  
\end{itemize}

Section II provides a brief overview of the important ideas related to point processes, 
collective coordinates, and hyperuniformity.  We apply these concepts in Section III 
to discuss how control over collective coordinates can be used to numerically 
generate configurations of hyperuniform point patterns, including those with anomalous 
asymptotic local density fluctuations, to a high numerical precision.  Section IV explores 
how increasing the fraction of constrained degrees of freedom within 
hyperuniform systems affects the observed pair correlations and, therefore, the local 
coordination structure.  In Section V we provide explicit calculations for the void statistics 
of our hyperuniform point patterns under weak and 
strong constraints, and we draw explicit connections among the regularity of the local 
structure, the exponential form of the small-wavenumber region of the structure factor, 
and the distribution of the local voids.  Concluding remarks are given in Section VI.

\section{Stochastic point patterns, collective coordinates, and hyperuniformity}

We consider many-particle configurations to be realizations of stochastic point processes in 
some subset of Euclidean space $\mathbb{R}^d$.  
A (finite) \emph{stochastic point pattern} is formally defined as a distribution of $N$ points 
$\{\mathbf{r}^N\}$ in some compact space $\mathcal{V}$ of volume (Lebesgue measure) $V$.  
We consider the case where the distribution is statistically homogeneous with periodic boundary 
conditions on $\mathcal{V}$; the thermodynamic limit $N, V\rightarrow +\infty$ with $\rho = N/V =$ 
constant can be taken appropriately to extend the point pattern to Euclidean 
space $\mathbb{R}^d$.  The statistics of the process are determined by an $N$-particle 
probability density function $P_N(\mathbf{r}^N)$, which need not be a Gibbs measure. 
Equivalently, one can specify the countable set of \emph{generic $n$-particle probability 
density functions} $\rho_n(\mathbf{r}^n)$, defined by
\begin{equation}
\rho_n(\mathbf{r}^n) = \frac{N!}{(N-n)!} \int P_N(\mathbf{r}^n, \mathbf{r}^{N-n}) d\mathbf{r}^{N-n}.
\end{equation}
The function $\rho_n$ is therefore the probability density associated with finding a 
subset of \emph{any} $n$ particles within volume elements $d\mathbf{r}^n$.  Note that 
for statistically homogeneous point patterns $\rho_1 = \rho$.  Related to the generic $n$-particle 
probability density function is the $n$-particle correlation function $g_n(\mathbf{r}^n)$, 
defined by
\begin{equation}
\rho^n g_n(\mathbf{r}^n) = \rho_n(\mathbf{r}^n).
\end{equation}
Of particular importance is the \emph{pair correlation function} $g_2(\mathbf{r})$, 
which can be made integrable by subtracting its long-range value of unity to give 
the \emph{total correlation function} $h(\mathbf{r}) = g_2(\mathbf{r}) - 1$.  A Fourier 
representation of $g_2(\mathbf{r})$ is given by the \emph{structure factor} $S(\mathbf{k})$, 
defined by
\begin{equation}
S(\mathbf{k}) = 1+\rho \hat{h}(\mathbf{k}),
\end{equation}
where we utilize the following convention for the Fourier transform:
\begin{equation}
\hat{f}(\mathbf{k}) = \int_{\mathbb{R}^d} \exp(-i\mathbf{k}\cdot\mathbf{r}) f(\mathbf{r}) d\mathbf{r}.
\end{equation}

Corresponding to any \emph{single} configuration of points $\{\mathbf{r}^N\}$ is a local 
density variable
\begin{equation}
\rho(\mathbf{r}) = \sum_{j=1}^N \delta(\mathbf{r}-\mathbf{r}_j),
\end{equation}
where $\delta$ denotes the Dirac delta function.  The ensemble average of this local 
density with respect to the statistics of the point process is
\begin{equation}
\langle \rho(\mathbf{r})\rangle = \rho,
\end{equation}
and the autocorrelation function is given by
\begin{equation}\label{seven}
\langle \rho(\mathbf{r}_1) \rho(\mathbf{r}_2)\rangle = \rho \delta(\mathbf{r}) + \rho^2 g_2(\mathbf{r})
\end{equation}
with $\mathbf{r} = \mathbf{r}_1 - \mathbf{r}_2$.  Note from \eqref{seven} that the 
autocorrelation function contains two contributions:  a delta function corresponding to 
the self-correlation of a point in the process and the pair correlation function between 
two distinct particles.  The self-correlation contribution is \emph{independent} of the 
distribution of particles in the system and arises for all correlated and uncorrelated point patterns.  

For statistically homogeneous point patterns subject to periodic boundary conditions, 
it is convenient to assume ergodicity and equate ensemble averages with volume 
averages over the unit cell.  This assumption is expected to be valid in the thermodynamic 
limit.  One can show that the volume-averaged local density and autocorrelation function are
\begin{align}
\overline{\rho(\mathbf{r})} &= \rho\label{eight}\\
\overline{\rho(\mathbf{x}+\mathbf{r})\rho(\mathbf{x})} &= \rho \delta(\mathbf{r}) + \frac{1}{V} 
\sum_{j\neq \ell = 1}^N \delta(\mathbf{r}-\mathbf{r}_{j\ell})\label{nine},
\end{align}
where $\mathbf{r}_{j\ell} = \mathbf{r}_j - \mathbf{r}_{\ell}$.  Equation \eqref{nine} suggests 
the following alternative definition of the pair correlation function:
\begin{equation}
\rho^2 g_2(\mathbf{r}) = \left\langle \frac{1}{V} \sum_{j\neq\ell = 1}^N \delta(\mathbf{r} 
- \mathbf{r}_{j\ell})\right\rangle.
\end{equation}
Since the Dirac delta functions in \eqref{nine} are by definition localized, this result has 
little practical utility when handling finite particle distributions.  However, one can take 
advantage of the periodicity of the unit cell to expand the local density in a Fourier series 
according to
\begin{equation}\label{eleven}
\rho(\mathbf{r}) = \frac{1}{V} \sum_{j=1}^N \sum_{\mathbf{k}} \exp\left[i \mathbf{k}
\cdot(\mathbf{r}-\mathbf{r}_j)\right],
\end{equation}
which is equivalent to a discrete (inverse) Fourier transform.  The wavevectors $\mathbf{k}$ 
in \eqref{eleven} are determined by the geometry of the unit cell; if the unit cell is formed with 
basis vectors $\{\mathbf{e}_i\}$, then the wavevectors satisfy
\begin{equation}
\mathbf{k}\cdot \mathbf{e}_i = 2\pi m
\end{equation}
for all $i$ and for some $m \in \mathbb{Z}$.  For simplicity, we will henceforth consider a 
$d$-dimensional cubic cell $[0, L]^d \subset \mathbb{R}^d$, which implies $\mathbf{k} 
= 2\pi \mathbf{n}/L$ for some $\mathbf{n} \in \mathbb{Z}^d$.  
Rewriting \eqref{eleven} in the form
\begin{equation}
\rho(\mathbf{r}) = \frac{1}{V} \sum_{\mathbf{k}} \exp(i \mathbf{k}\cdot \mathbf{r}) 
\hat{\rho}(\mathbf{k}),
\end{equation}
where
\begin{equation}\label{fourteen}
\hat{\rho}(\mathbf{k}) = \sum_{j=1}^N \exp(-i\mathbf{k}\cdot\mathbf{r}_j),
\end{equation}
we observe that the local density is the discrete (inverse) Fourier transform of 
$\hat{\rho}$, which we call a \emph{collective density variable}.  

The identity \eqref{eight} can also be obtained using the Fourier representation 
\eqref{eleven}, meaning that only the mode $\mathbf{k} = \mathbf{0}$ contributes to 
the local density on average.
However, the autocorrelation function is now of the form
\begin{align}
\overline{\rho(\mathbf{x}+\mathbf{r}) \rho(\mathbf{x})} &= \frac{\rho}{V} \sum_{\mathbf{k}} 
\exp(i \mathbf{k}\cdot\mathbf{r}) + \frac{1}{V^2} \sum_{j\neq\ell} \sum_{\mathbf{k}} 
\exp\left[i \mathbf{k}\cdot (\mathbf{r}-\mathbf{r}_{j\ell})\right]\\
&= \rho \delta(\mathbf{r}) + \frac{\rho^2}{N^2} \sum_{\mathbf{k}} \exp(i \mathbf{k}
\cdot\mathbf{r})\left[\lvert \hat{\rho}(\mathbf{k})\rvert^2 - N\right],
\end{align}
which, by comparing with \eqref{seven}, implies \cite{FN1}
\begin{equation}\label{seventeen}
g_2(\mathbf{r}) = \frac{1}{N^2} \sum_{\mathbf{k}} \exp(i \mathbf{k}\cdot \mathbf{r}) 
\left[\lvert\hat{\rho}(\mathbf{k})\rvert^2 - N\right].
\end{equation}
The result \eqref{seventeen} allows one to directly compute the pair correlation 
function from the collective density variables $\hat{\rho}$; note that the $\mathbf{k} 
= \mathbf{0}$ mode must be included in this calculation to ensure the correct long-range 
behavior $g_2(\mathbf{r}) \rightarrow 1$ as $\lVert\mathbf{r}\rVert \rightarrow +\infty$.  
In practice, one must truncate the wavevector summation in \eqref{seventeen}, leading to 
oscillatory approximations to $g_2$ within some threshold determined by the cut-off 
magnitude of the wavevectors.   

\emph{Hyperuniform} point patterns constitute a subclass of point processes lacking 
infinite-wavelength local density fluctuations \cite{ToSt03}.  Specifically, it has been 
shown that the variance $\sigma^2_N(R)$ in the number of points within a local 
spherical observation window $\mathcal{W}(R)$ of radius $R$ and volume $v(R)$ 
scales asymptotically as \cite{ToSt03}
\begin{equation}\label{hypscale}
\sigma^2_N(R) = \langle N(R)\rangle \left[ A_N(R) +  B_N(R)/R + \text{lower-order terms}\right],
\end{equation}
where $\langle N(R)\rangle = \rho v(R)$ is the average number of points in the observation window.  The coefficients $A_N(R)$ and $B_N(R)$ in \eqref{hypscale} are determined solely by the two-particle information of the point pattern:
\begin{align}
A_N(R) &= 1+ \rho \int_{\mathcal{W}(R)} h(\mathbf{r}) d\mathbf{r} \qquad (R\rightarrow +\infty)\\
B_N(R) &= -\frac{\rho \Gamma(1+d/2)}{\Gamma[(d+1)/2]\Gamma(1/2)} 
\int_{\mathcal{W}(R)} h(\mathbf{r}) r d\mathbf{r} \qquad (R\rightarrow +\infty).
\end{align}
So long as $h(r) \rightarrow 0$ faster than $r^{-d}$, the leading-order 
coefficient $A_N(R)$ converges asymptotically as $A_N(R) = A_N \equiv 
\lim_{\lVert\mathbf{k}\rVert\rightarrow 0}S(\mathbf{k})$ \cite{FN2}.  
By definition, a hyperuniform point pattern possesses a number variance growing slower 
than the volume $v(R)$ of the observation window (equivalently, the mean 
number of points $\langle N(R)\rangle$), implying that $A_N = 0$ and infinite-wavelength 
density fluctuations vanish.  

The most common examples, including all Bravais lattices, periodic non-Bravais lattices, 
quasicrystals possessing Bragg peaks, and certain disordered point patterns with 
pair correlation functions decaying to unity exponentially fast, of hyperuniform point patterns 
possess constant number variance coefficients $B_N(R) = B_N$ \cite{ToSt03}.  
This behavior implies that the isotropic structure factor $S(k)$ possesses a small-wavenumber 
scaling $Dk^{\alpha}$ with $\alpha \geq 2$, including the special case 
$\alpha = +\infty$ for periodic structures.  However, it is also possible to find 
hyperuniform point patterns for which $0 < \alpha < 2$, in which case 
$C_1 \leq B_N(R) \leq C_2 R$ as $R\rightarrow +\infty$ for some constants 
$C_1$ and $C_2$.  The most well-known examples of these types of 
``anomalous'' local density fluctuations occur when $S(k) \sim k$ as $k\rightarrow 0$,
 in which case $B_N(R) = A_1 \ln(R) + A_2$ with $A_1$ and $A_2$ constant.  
 This situation has been well-characterized in three-dimensional maximally random 
jammed packings of hard spheres \cite{DoStTo05}, the ground states of liquid
 helium \cite{ReCh67}, and noninteracting spin-polarized fermion ground states \cite{ToScZa08}.    
 However, examples where $\alpha < 1$ have heretofore not appeared in the 
literature.

\section{Collective coordinate construction of hyperuniform point patterns}

One goal of this work is to construct examples of hyperuniform point patterns 
possessing the aforementioned ``anomalous'' asymptotic local density fluctuations, 
meaning that the number variance grows slower than the volume of an observation 
window but faster than the surface area.  Collective density variables provide an 
attractive means to control the small-wavenumber region of the structure factor 
$S(\mathbf{k})$, thereby allowing us to construct a hyperuniform point pattern 
with targeted local density fluctuations.  Specifically, we define an objective 
function $\Phi$ according to
\begin{equation}\label{eighteen}
\Phi(\mathbf{r}^N) = \sum_{\mathbf{k} \in \mathcal{Q}} \left[S(\mathbf{k}; 
\mathbf{r}^N)-S_0(\mathbf{k})\right]^2,
\end{equation}
where $S_0(\mathbf{k})$ is the targeted form of the structure factor and 
$\mathcal{Q}$ denotes some finite subset of wavevectors $\mathbf{k}$.  
The structure factor is determined using collective density variables; specifically,
\begin{equation}\label{nineteen}
S(\mathbf{k}; \mathbf{r}^N) = \frac{\lvert\hat{\rho}(\mathbf{k})\rvert^2}{N} 
\qquad (\mathbf{k}\neq \mathbf{0}),
\end{equation}
where $\hat{\rho}(\mathbf{k})$, implicitly a function of the particle positions 
$\mathbf{r}^N$, is defined by \eqref{fourteen}.  
The zero-wavevector is excluded from \eqref{nineteen} since it provides an 
$\mathcal{O}(N)$ contribution to the structure factor, corresponding to a delta 
function in the thermodynamic limit from the long-range behavior of $g_2$.
By expanding \eqref{eighteen}, one can show that our minimization problem 
corresponds to finding the classical 
ground state of a many-particle system with up to four-body interactions \cite{UcToSt06}
\begin{equation}
\Phi(\mathbf{r}^N) = \sum_{i\neq j\neq \ell \neq m} v_4(\mathbf{r}_i, \mathbf{r}_j, 
\mathbf{r}_\ell, \mathbf{r}_m) + \sum_{i\neq j \neq \ell} v_3(\mathbf{r}_i, \mathbf{r}_j, 
\mathbf{r}_\ell) + \sum_{i\neq j} v_2(\mathbf{r}_i, \mathbf{r}_j) + v_0,
\end{equation}
where
\begin{align}
v_4(\mathbf{r}_i, \mathbf{r}_j, \mathbf{r}_\ell, \mathbf{r}_m) &= \frac{1}{N^2}
\sum_{\mathbf{k}\in \mathcal{Q}} \cos(\mathbf{k}\cdot\mathbf{r}_{ij})
\cos(\mathbf{k}\cdot\mathbf{r}_{\ell m})\\
v_3(\mathbf{r}_i, \mathbf{r}_j, \mathbf{r}_\ell) &= \frac{4}{N^2} \sum_{\mathbf{k}
\in \mathcal{Q}} \cos(\mathbf{k}\cdot\mathbf{r}_{ij})\cos(\mathbf{k}\cdot\mathbf{r}_{i\ell})\\
v_2(\mathbf{r}_i, \mathbf{r}_j) &= \frac{2}{N}\sum_{\mathbf{k}\in \mathcal{Q}} 
\cos(\mathbf{k}\cdot\mathbf{r}_{ij})[1-S_0(\mathbf{k})]\\
v_0 &= \sum_{\mathbf{k}\in\mathcal{Q}} \left[S_0(\mathbf{k}) - 1\right]^2.
\end{align}

The set $\mathcal{Q}$ in \eqref{eighteen} is chosen to contain all wavevectors, 
excluding the zero mode, with norm less than some upper bound $K$.  
This construction allows us to target specifically the small-wavenumber region 
of the structure factor, which controls the asymptotic local density fluctuations.  
The target function $S_0$ is chosen with the form
%\begin{equation}
%S_0(\mathbf{k}) = D\lVert\mathbf{k}\rVert^{\alpha),
%\end{equation}
\begin{equation}\label{twenty}
S_0(\mathbf{k}) = D\lVert\mathbf{k}\rVert^{\alpha} \qquad \text{for all } \mathbf{k}\in \mathcal{Q}.
\end{equation}
In order for the target function to correspond to a realizable point pattern, 
it is necessary that $D \geq 0 $ to enforce positivity of the structure factor.  
The parameter $\alpha$ determines the asymptotic behaviors of the pair 
correlation function and the number variance [c.f. \eqref{NVscaling}].  
Previous work \cite{UcToSt06}  has considered the cases $\alpha = 1, 2, 4, 6, 8,$ and $10$ 
in dimensions $d = 2$ and $3$.  
It has recently been conjectured that $\alpha = 1$ corresponds to the \emph{minimal} 
exponent consistent with the constraints of saturation and strict jamming in sphere 
packings \cite{ZaJiTo10}; however, 
systems for which $\alpha < 1$ have not been reported in the literature, and their 
statistical properties are unknown.  
The objective function \eqref{eighteen} is minimized to within $10^{-17}$ of its global 
minimum using the MINOP algorithm \cite{DeMe79, Ka99}, which has several 
computational advantages 
for this type of investigation as previously reported in the literature \cite{UcToSt06}.  
MINOP applies a dogleg strategy that uses a gradient direction when one is far 
from the minimum, a quasi-Newton direction when one is close, and a linear 
combination of the two when one is at intermediate distances from the minimum.  

It is important for this study to verify that the constructed point patterns are indeed 
hyperuniform with the correct targeted asymptotic local density fluctuations.  
This criterion requires high resolution of the small-wavenumber region of the structure 
factor.  Specifically, the smallest observable wavenumber magnitude in the collective 
coordinates representation 
(in a $d$-dimensional cubic unit cell) is $k_{\text{min}} = 2\pi/L = 2\pi \rho^{1/d}/N^{1/d}$, 
where $L$ is the box length, $N$ is the number of particles, and $\rho$ is the number density.  
To ensure hyperuniformity, we therefore require that $\lim_{N\rightarrow +\infty} 
S(k_{\text{min}}) = 0$, where the limit is taken at constant density.  

Since any simulation necessarily requires choosing $N$ finite, it is essential to select a 
value of $N$ sufficiently large to enforce both hyperuniformity and the desired form of 
the structure factor near the origin.  
Unfortunately, the $\mathcal{O}(N^{-1/d})$ scaling of $k_\text{min}$ makes obtaining 
such resolution increasingly difficult in higher dimensions.  
Our interest is in verifying the existence of anomalous hyperuniform point patterns 
and understanding their statistical properties, and we therefore limit our studies to 
one dimension, where the scaling is most favorable, with $N = 2000$ particles.  
It should be appreciated, however, that hyperuniform point patterns with 
logarithmically-growing asymptotic density fluctuations are known in arbitrarily 
high dimensions \cite{ToScZa08}.  
Importantly, since our minimization procedure is equivalent to finding the 
classical ground state of a long-range interaction with up to four-body 
potentials and can be used in principle to construct hyperuniform point patterns 
in any dimension,
nontrivial phase behaviors can still be observed \cite{LiMa66}, and we are therefore 
able to extend our conclusions to higher-dimensional structures.

\section{Collective coordinates and realizability of point patterns}

For a general $d$-dimensional point pattern of $N$ particles, there are $dN$ translational 
degrees of freedom in the absence of constraints on the system.  One must therefore 
choose a set of wavevectors $\mathcal{Q}$ 
for the objective function \eqref{eighteen} containing only a fraction $\chi$ of these 
degrees of freedom.  In one dimension there are $2M(K) = \text{floor}(KL/\pi)$ wavevectors, 
excluding the zero mode, with magnitude less than or equal to $K$.  
Inversion-invariance of the modulus of the collective density variable 
implies that $M(K)$ of these wavevectors can be independently constrained; 
we therefore define a new parameter
\begin{equation}\label{chidef}
\chi = \frac{M(K)}{dN},
\end{equation}
which represents the fraction of independently constrained degrees of 
freedom from the objective function $\Phi$.  

For the case where the targeted structure factor $S_0(\mathbf{k}) = 0$ 
for all $\mathbf{k} \in \mathcal{Q}$, it has been previously shown \cite{FaPeStSt91} 
that increasing the parameter $\chi$ induces a greater degree of order on the particle distribution.  
Specifically, in one dimension the corresponding point patterns are disordered 
for $0 < \chi < 1/3$ and crystalline for $\chi > 1/2$ \cite{FN3};  intermediate 
values of $\chi$ interpolate between these two regimes \cite{FNhighd}.  
However, it is known that target functions of the form \eqref{twenty} interfere 
with this order-disorder phase transition; here we provide analytic results 
suggesting that this transition is shifted to 
higher values of $\chi$ for all finite $\alpha$ \cite{FN4}.

For a one-dimensional point pattern, the wavevectors are of the form $k = 2\pi m/L$ 
for $m \in \mathbb{Z}$, and one can write the collective density variable as:
\begin{equation}
\hat{\rho}(m) = \sum_{j = 1}^N \exp(-i 2\pi m r_j/L).
\end{equation}
Additionally, the total correlation function is of the form [cf. \eqref{seventeen}]
\begin{equation}\label{twentythree}
h(r) = \frac{2}{N^2}\sum_{m=1}^{+\infty} \cos(2\pi m r/L)\left[\lvert\hat{\rho}(2\pi m/L)\rvert^2 
- N\right],
\end{equation}
which for the targeted point pattern can be decomposed as:
\begin{align}
h(r) &= \frac{2}{N} \sum_{m=1}^M \cos(2\pi m r/L)\left[D (2\pi m/L)^{\alpha} - 1\right] 
+ \frac{2}{N^2} \sum_{m = M+1}^{+\infty} \cos(2\pi m r/L) \left[\lvert\hat{\rho}(2\pi m/L)\rvert^2 
- N\right]\\
&= h_0(r; M) + h_1(r; M),
\end{align}
where $h_0(r; M)$ is the contribution to the total correlation function due to 
\emph{constrained} wavevectors and $h_1(r; M)$ is the \emph{unconstrained} 
contribution.  The function $h_0$ can be simplified as
\begin{align}
h_0(r; M) &= \left(\frac{2^{\alpha+1} \pi^{\alpha}D}{N L^{\alpha}}\right) 
\sum_{m=1}^M \cos(2\pi m r/L) m^{\alpha} - \frac{2}{N} \sum_{m=1}^M \cos(2\pi m r/L)\\
&= C(\alpha, D) \sum_{m=1}^M \cos(2\pi m r/L) m^{\alpha} - (2/N) \cos[(M+1)\pi r/L]
\csc(\pi r/L) \sin(M\pi r/L)
\end{align}
where
\begin{equation}
C(\alpha, D) = \frac{2^{\alpha+1} \pi^{\alpha} D}{N L^{\alpha}}
\end{equation}
is a parameter-dependent constant.  %Figure [REF]  plots $h_0(r; M)$ for increasing values of $M$.
%\begin{figure}[!tp]
%\centering
%\includegraphics[width=0.75\textwidth]{g2_constr}
%\caption{Constrained contribution $h_0(r; M)+1$ to the pair correlation function $g_2(r)$ for increasing values of $M$.}
%\end{figure}
The global minimum of $h_0(r; M)$ occurs at $r = 0$, corresponding to
\begin{align}
h_0(0; M) &= C(\alpha, D) \sum_{m=1}^M m^{\alpha} - (2M/N)\\
&= C(\alpha, D) H^{(-\alpha)}(M) - (2M/N)\label{thirty},
\end{align}
where 
\begin{equation}
H^{(\alpha)}(n) = \sum_{m=1}^n m^{-\alpha}
\end{equation}
is the \emph{harmonic number} of order $\alpha$.

The negative contribution to $h_0(0; M)$ in \eqref{thirty} suggests that 
there may be an upper threshold $M^*$ beyond which $h_0(0; M) < 0$.  
For any values of $M$ in this region, the constrained contribution $h_0$ 
to the total correlation function of the point pattern is no longer in 
itself \emph{realizable} as a point process.  The realizability problem in 
classical statistical mechanics \cite{ToSt02} and the associated $N$-representability 
problem in quantum statistics \cite{Co63} are notoriously difficult and unsolved 
problems in physics that ask under what sufficient and necessary conditions a 
reduced two-particle correlation function can be expressed as the integral over 
a full $N$-particle probability density.  In the classical case, one can consider 
specifying a pair correlation function $g_2$ and attempting to construct a 
corresponding point process.  Known necessary realizability conditions on $g_2$ include
\begin{align}
g_2(\mathbf{r}) &\geq 0 \text{ for all } \mathbf{r}\\
S(\mathbf{k}) &\geq 0 \text{ for all } \mathbf{k}
\end{align}
along with the somewhat weaker Yamada condition 
\begin{equation}
\sigma^2_N(R) \geq \theta (1-\theta)
\end{equation}
on the fractional part $\theta$ of the average number of particles in an observation 
window \cite{Ya61}.  
The Yamada condition appears easy to satisfy in all but relatively low dimensions \cite{ToSt02}.
The determination of other realizability conditions on $g_2$ is an open problem \cite{KuLeSp07}.  

\begin{figure}[!tp]
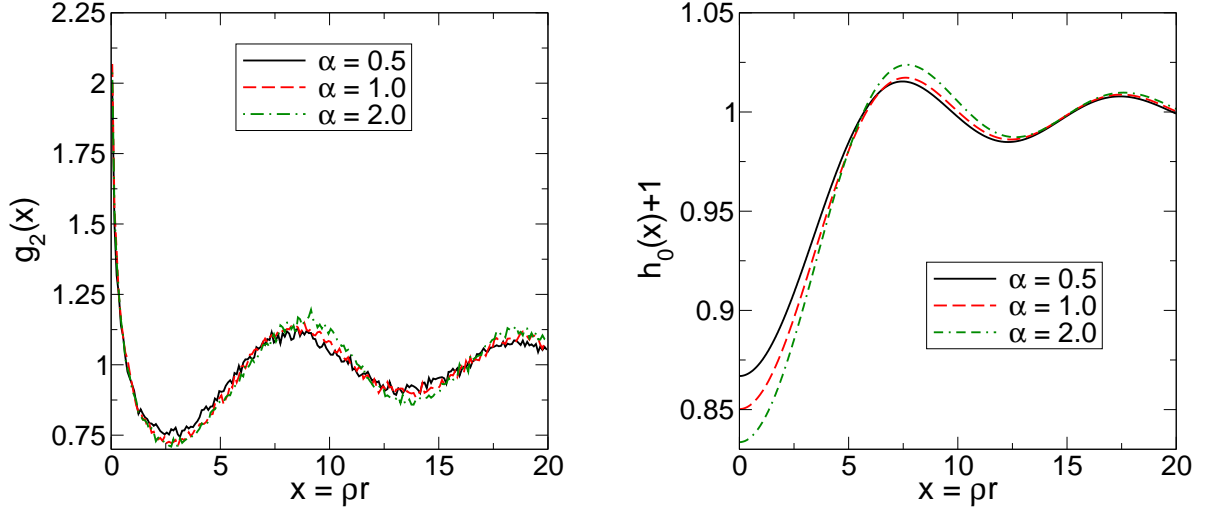

\centering
\includegraphics[width=0.45\textwidth]{Fig2A}\hspace{0.05\textwidth}
\includegraphics[width=0.45\textwidth]{Fig2B}
\caption{(Color online)  Left panel:  Pair correlation function $g_2$ for 
numerically-constructed hyperuniform point patterns with small-wavenumber 
scalings $Dk^{\alpha}$ and $\chi = 0.1$.  Right panel:  Constrained contributions 
to the pair correlation functions.}\label{chi01g2}
\end{figure}
\begin{figure}[!tp]
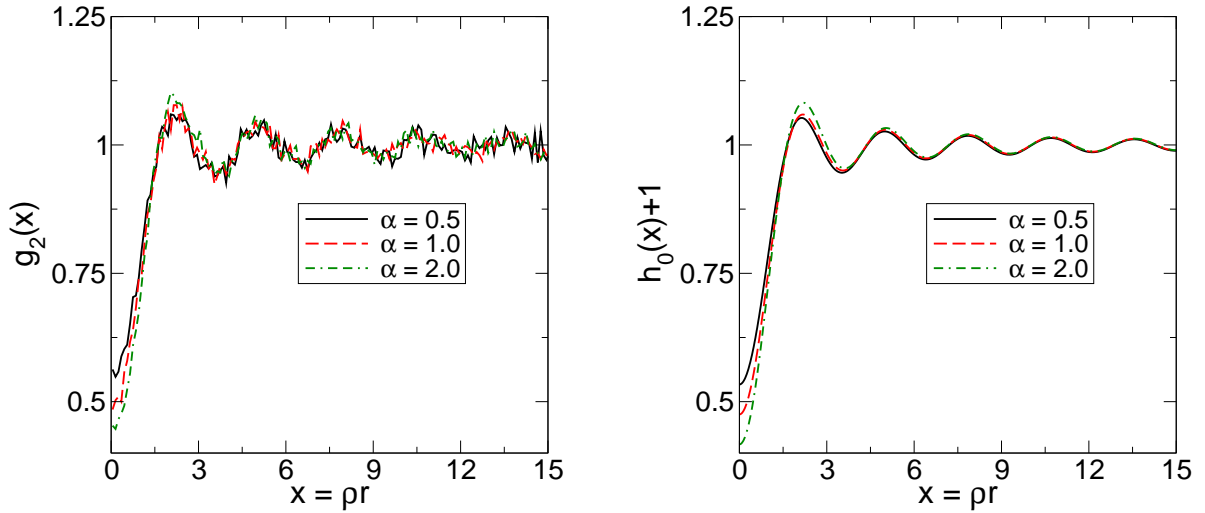

\centering
\includegraphics[width=0.45\textwidth]{Fig3A}\hspace{0.05\textwidth}
\includegraphics[width=0.45\textwidth]{Fig3B}
\caption{(Color online)  Left panel:  Pair correlation function $g_2$ for 
numerically-constructed hyperuniform point patterns with small-wavenumber 
scalings $Dk^{\alpha}$ and $\chi = 0.35$.  Right panel:  Constrained contributions 
to the pair correlation functions.}\label{chi035g2}
\end{figure}
Figures \ref{chi01g2} and \ref{chi035g2} compare the pair correlation functions 
and the constrained contributions $h_0(r) + 1$ for numerically-constructed 
point patterns (using the methodology of Section III)
with small-wavenumber exponents $\alpha = 0.5, 1.0,$ and $2.0$ and $\chi = 0.1$ and $0.35$.    
For $\chi = 0.1$, corresponding to a small fraction of constrained degrees of freedom, 
the constrained contribution $h_0(r) + 1$ places only moderate constraints on the 
local structure of the system, primarily controlling oscillations in $g_2$ beyond approximately 
five nearest-neighbor distances.  Interestingly, the small-$r$ behaviors of $g_2(r)$ 
and $h_0(r)+1$ are strikingly different.  Although the \emph{constrained} contribution 
to the pair correlation function generates an effective repulsion between particle pairs, 
the full pair correlation function indicates a tendency for particles to cluster at short pair 
separations.  It follows that the unconstrained contribution to the pair correlation function 
plays a substantial role in determining the local structure for this 
system.    

\begin{figure}[!t]
\centering
\includegraphics[width=0.5\textwidth]{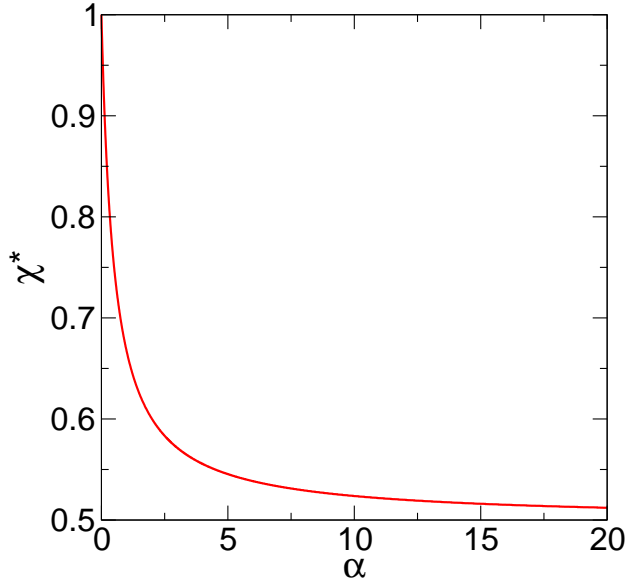}
\caption{Typical threshold values $\chi^*(\alpha)$ beyond which the \emph{constrained} 
contribution $h_0(r)$ to the total correlation function is no longer \emph{realizable} as a 
point process.  This curve corresponds to choosing $S(K) = 0.5,$ where $K$ is the 
magnitude of the maximally constrained wavevector.  Note that as $\alpha \rightarrow +\infty$ 
we recover the crystallization threshold  $\chi^* = 0.5$ 
reported in Ref. \cite{FaPeStSt91}.}\label{achi}
\end{figure}
However, the situation is quite different upon increasing the 
constrained degrees of freedom to $\chi = 0.35$.  Figure \ref{chi035g2} 
shows that the constrained contribution to $g_2$ almost exactly mirrors 
the full pair correlation function, implying that sufficiently constraining the 
collective density variables places a \emph{strong} constraint on the local 
structure of the point pattern.  It follows that the value $M^*$ beyond which 
$h_0(0; M) < 0$ is an indicative precursor to the \emph{loss of realizability} of 
the targeted structure factor.  We have mapped the threshold value $M^*$ 
(equivalently, $\chi^*$) in Fig. \ref{achi}.  We emphasize that this loss of 
realizability is associated with negativity of the real-space pair correlation 
function; the structure factor itself is still positive over its entire domain.    
Interestingly, as the exponent $\alpha$ controlling the small-wavenumber 
region of the structure factor increases, we recover the value $\chi = 0.5$ 
corresponding to crystallization in the case where $S_0(\mathbf{k}) = 0$ 
for all $\mathbf{k} \in \mathcal{Q}$.    This observation suggests that the 
threshold values of $\chi$ beyond which $h_0(0; M) < 0$ generalize this phase transition.
In Section V, we provide additional arguments to support this claim.  

\section{Void statistics and coordination structure}

\subsection{Exclusion probability functions}

The $n$-particle correlation functions contain information concerning the 
relative locations of points within a point process, and, in principle, specifying 
the countably infinite set (in the thermodynamic limit) of such functions is 
sufficient to completely determine the point pattern.  However, any finite 
collection of correlation functions contains only partial details of the spatial 
arrangements of the points, implying that there are degenerate structures 
with these same statistics \cite{JiStTo10}.  In particular, the $n$-particle 
correlations functions do not in themselves provide direct information about 
the space \emph{exterior} to the points, or the so-called \emph{void space}.    
It has been shown for point patterns \cite{ZaJiTo10, ToLuRu90} 
(and random media \cite{Tobook}) that the distribution of the void space 
is indeed a more fundamental descriptor of the point process than the 
arrangements of the points themselves.   
Here we are interested in characterizing the relationship between asymptotic 
local number density fluctuations and the void space statistics; in particular, we 
would like to examine the constraints that the exponent $\alpha$ in the 
small-wavenumber region of the structure factor places on the distribution of the void space.  

\begin{figure}[!t]
\centering
\includegraphics[width=0.5\textwidth]{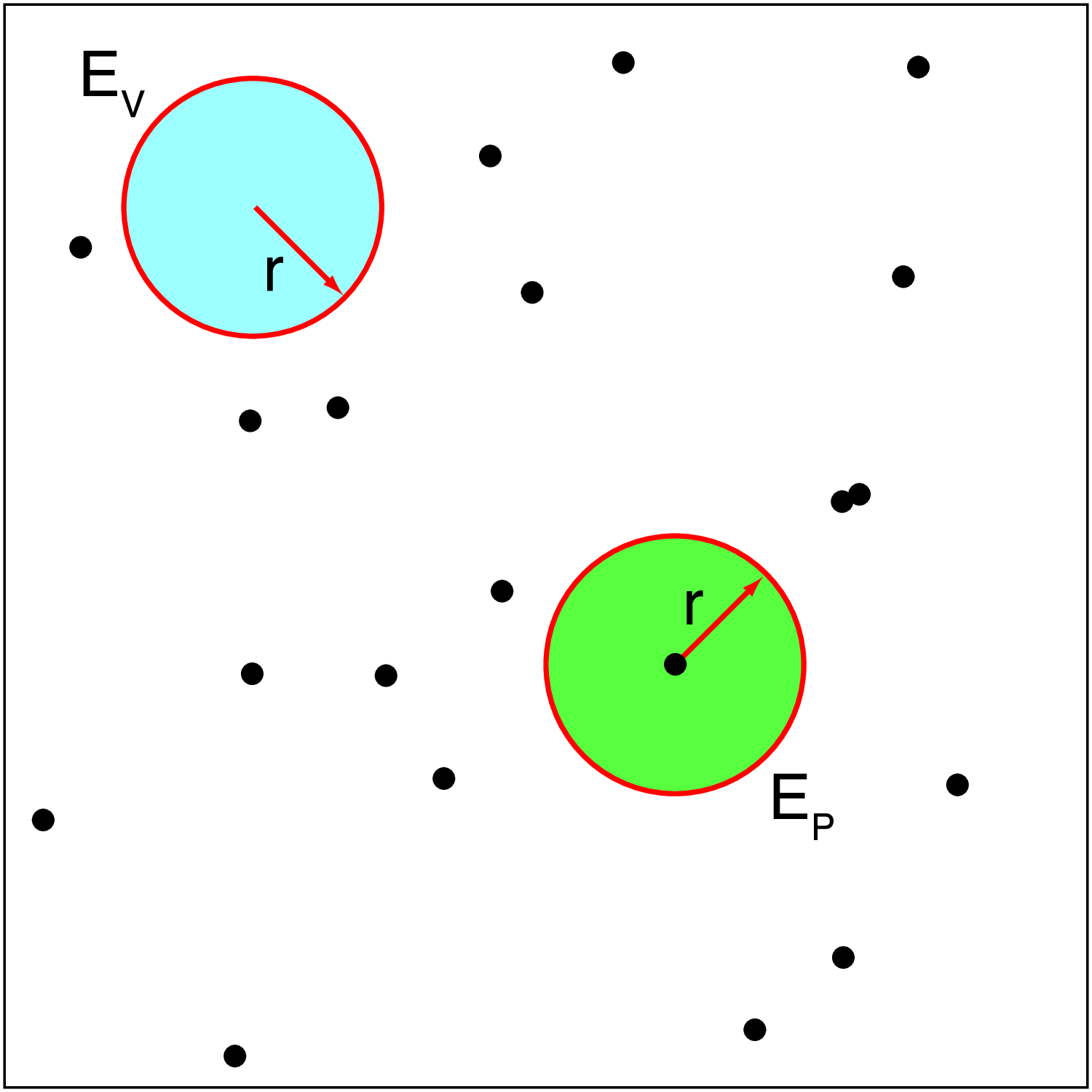}
\caption{Events contributing to the void exclusion probability $E_V(r)$ (upper left) and the 
particle exclusion probability $E_P(r)$ (lower).  The points correspond to a realization of a 
disordered point process.}\label{E}
\end{figure}
One can define two types of ``exclusion'' functions, both of which measure the 
availability of empty space surrounding points of a stochastic process.  
The \emph{void exclusion probability function} $E_V(r)$ is the probability of 
finding a $d$-dimensional spherical cavity of radius $r$ centered at an arbitrary 
position in $\mathbb{R}^d$.  
The void exclusion probability has recently been shown to play a fundamental 
role in the covering and quantizer problems from discrete geometry and number 
theory \cite{To10}.
Closely related to this descriptor is the \emph{particle exclusion probability 
function} $E_P(r)$, which is the probability of finding a $d$-dimensional sphere 
of radius $r$ centered on a point of the point process but containing no other points.  
Figure \ref{E} highlights the differences between these functions.   

The exclusion probability functions are complementary cumulative distributions 
of the void and particle nearest-neighbor functions $H_V(r)$ and $H_P(r)$, 
respectively \cite{ToLuRu90, Tobook}.  
The void nearest-neighbor function is the probability density of a finding the 
nearest point of a point process with respect to an arbitrary location in 
$\mathbb{R}^d$ within a radial distance $r+dr$.  The particle nearest-neighbor 
function is defined similarly but with respect to nearest neighbors between two 
points of a point process.  One therefore has the following simple relationships 
between these sets of functions:
\begin{align}
H_V(r) &= -\frac{\partial E_V(r)}{\partial r}\\
H_P(r) &= -\frac{\partial E_P(r)}{\partial r}.
\end{align}

One can relate the void and particle exclusion probabilities via a simple 
probabilistic construction \cite{ScZaTo09}.  Specifically, we consider a 
generalized exclusion probability $E_V(r; \epsilon)$, which is the probability 
of finding a $d$-dimensional annulus of outer radius $r$ and inner radius 
$\epsilon$; by definition, $E_V(r; 0) = E_V(r)$.  Taking the derivative of this 
function with respect to the inner radius $\epsilon$ gives a function proportional 
to the probability of finding a point within a small radial region inside the annulus 
and the annulus itself devoid of points.  It follows that $E_P(r)$, the conditional 
probability of finding a spherical cavity centered on a point, is
\begin{equation}
E_P(r) = \lim_{\epsilon\rightarrow 0^+} \frac{1}{\rho s(\epsilon)} \frac{\partial 
E_V(r; \epsilon)}{\partial \epsilon},
\end{equation}
where $s(\epsilon)$ is the surface area of a $d$-dimensional sphere of 
radius $\epsilon$.  This construction is known in the theory of point 
processes \cite{DaVe08} and has also been used in the literature to 
identify the void statistics of certain point patterns related to problems 
in number theory, random matrix theory, and quantum mechanics \cite{ToScZa08}.  
One can without loss of generality define the \emph{exclusion correlation function} 
$\eta(r)$ according to 
\begin{equation}
\eta(r) \equiv \frac{E_P(r)}{E_V(r)}.
\end{equation}
This function provides a measure of the correlations between $neighboring$ 
points in a stochastic point pattern and is identically unity for a Poisson point process. 
It is interesting to note that for a system of \emph{equilibrium} hard spheres 
of diameter $D$, the 
exclusion correlation function is given by \cite{ToLuRu90}
\begin{equation}
\eta(r) = \begin{cases}
[E_V(r)]^{-1}, & r\leq D\\
[E_V(D)]^{-1}, & r\geq D,
\end{cases}
\end{equation}
which depends only on knowledge of $E_V(r)$ and is monotonically 
nondecreasing for all $r$ with $\eta(0) = 1$.  

Further insight into the probabilistic meanings of $E_P$, $E_V$, and $\eta$ 
can be gained by introducing the notion of the particle space, defined to be 
the subset (of Lesbesgue measure zero) of $\mathbb{R}^d$ occupied by the 
points of the point process.  The particle exclusion probability function $E_P(r)$ 
is then the fraction of the particle space that can be decorated by a $d$-dimensional 
sphere of radius $r$ containing no other points of the process.  To define the void 
exclusion probability $E_V(r)$, one decorates all of the points in the process by 
spheres of radius $r$ and then determines the fraction of \emph{all} space not 
occupied by the spheres; this value corresponds to the portion of space available 
to insert a cavity of radius $r$ \cite{To10}.
The exclusion correlation function $\eta(r)$ then provides a measure of the relative 
available space for a cavity of radius $r$ in the particle space compared to the external 
void space.  

Torquato and coworkers \cite{ToLuRu90} have provided the following series 
representations for the exclusion probability functions:
\begin{align}
E_V(r) &= 1+\sum_{k=1}^{+\infty} \frac{(-\rho)^k}{\Gamma(k+1)} \int g_k(\mathbf{r}^k) 
\prod_{j=1}^k m(\lVert \mathbf{x}-\mathbf{r}_j\rVert; r) d\mathbf{r}_j\\
E_P(r) &= 1+\sum_{k=1}^{+\infty} \frac{(-\rho)^k}{\Gamma(k+1)} \int g_{k+1}(\mathbf{r}^{k+1}) 
\prod_{j=2}^{k+1} m(\lVert\mathbf{r}_1-\mathbf{r}_j\rVert; r) d\mathbf{r}_j,
\end{align}
where $m(r; R) = \Theta(R-r)$.  Since these functions are special cases of a more general 
canonical $n$-particle correlation function \cite{To86}, one can establish rigorous upper 
and lower bounds by truncating these series at finite order.  Specifically, by writing
\begin{equation}
E_{V/P}(r) = \sum_{k=0}^{+\infty} E_{V/P}^{(k)}(r),
\end{equation}
where $E_{V/P}^{(0)} \equiv 1$, we have the following hierarchy of bounds:
\begin{align}
E_{V/P}(r) &\leq \sum_{k=0}^{\ell}E_{V/P}^{(k)}(r) \qquad (\ell \text{ even})\label{lbound}\\
E_{V/P}(r) &\geq \sum_{k=0}^{\ell} E_{V/P}^{(k)}(r) \qquad (\ell \text{ odd}),
\end{align}
 which become sharper with increasing $\ell$.

\subsection{Local statistics of anomalous hyperuniform point patterns}

\begin{figure}[!t]
\centering
\includegraphics[height=2.3in]{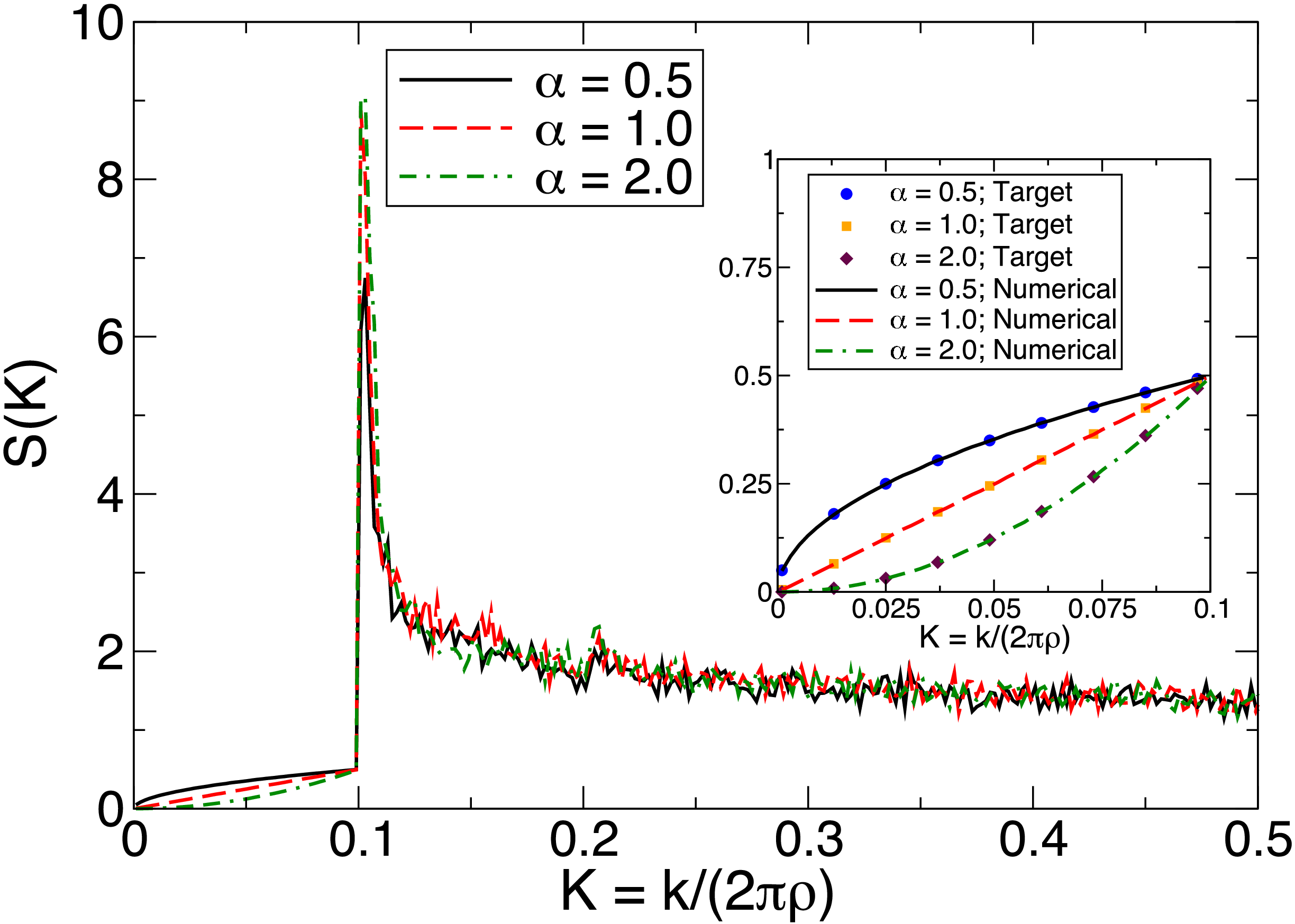}\hspace{0.05\textwidth}
\includegraphics[width=0.4\textwidth]{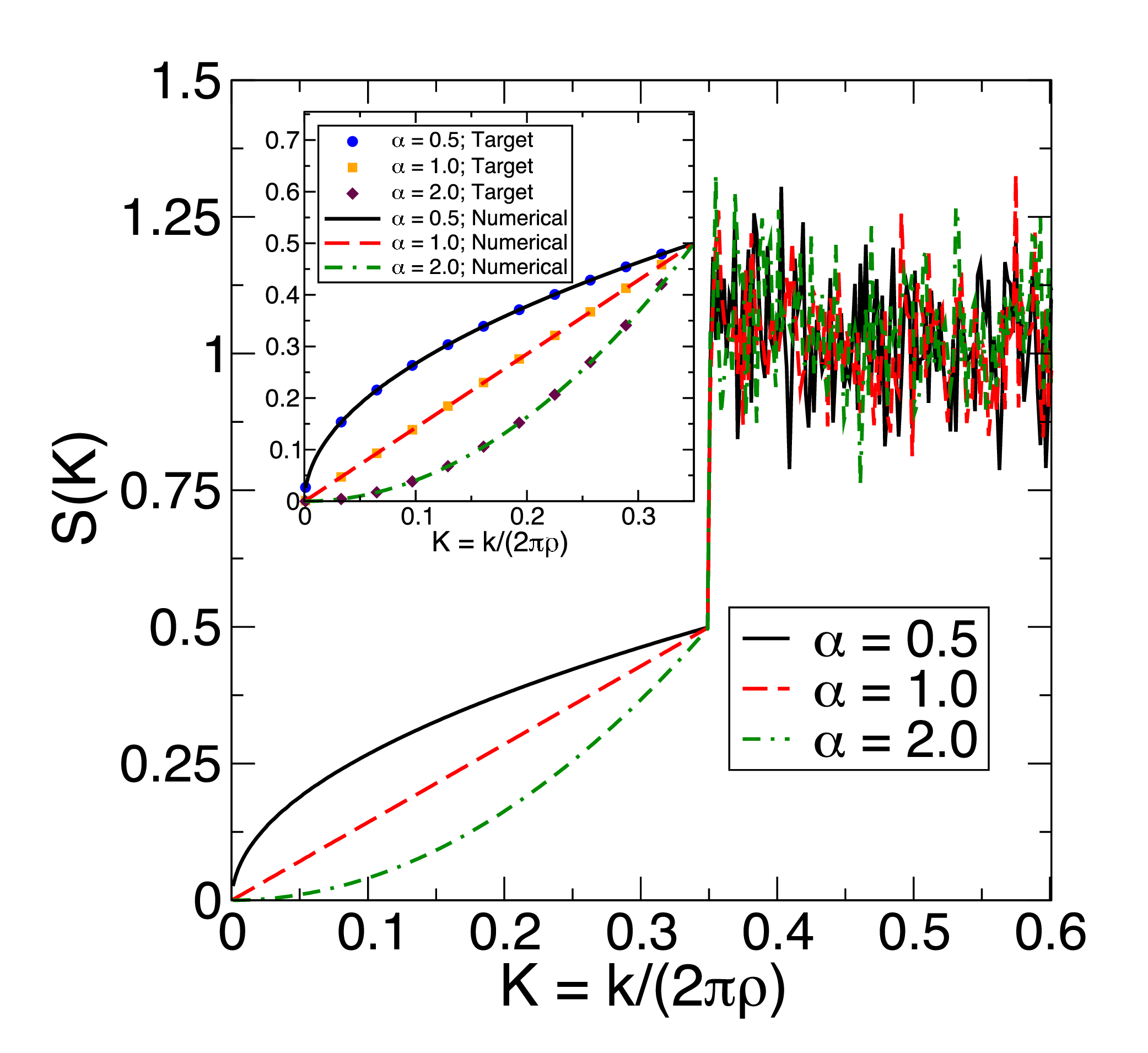}
\caption{(Color online)  Left panel:  Structure factor with small-$k$ behavior 
$Dk^{\alpha}$ (inset) for numerically-constructed hyperuniform point patterns 
with $\chi = 0.1$.  Right panel:  Structure factor with small-$k$ behavior for 
$\chi = 0.35$.}\label{Sk}
\end{figure}
We have been able to successfully construct point configurations exhibiting 
anomalous asymptotic local number density fluctuations.  Figure \ref{Sk} provides 
images of the structure factors for our configurations at $\chi = 0.1$ and $\chi=0.35$.  
As we show, for all wavevectors within the constrained portion of the spectrum, the 
structure factor matches its target value within an exceedingly small numerical 
tolerance (on the order of $10^{-17}$).  In addition to the systems shown, we have 
also been able to reliably construct configurations with small-$k$ exponential 
behaviors $\alpha \geq 0.25$.  In order to keep the exposition clear, we have only 
presented results for $\alpha = 0.5$ with the disclaimer that our conclusions will 
apply for other point patterns with anomalous local number density fluctuations.

\begin{figure}[!t]
\centering
\includegraphics[width=0.45\textwidth]{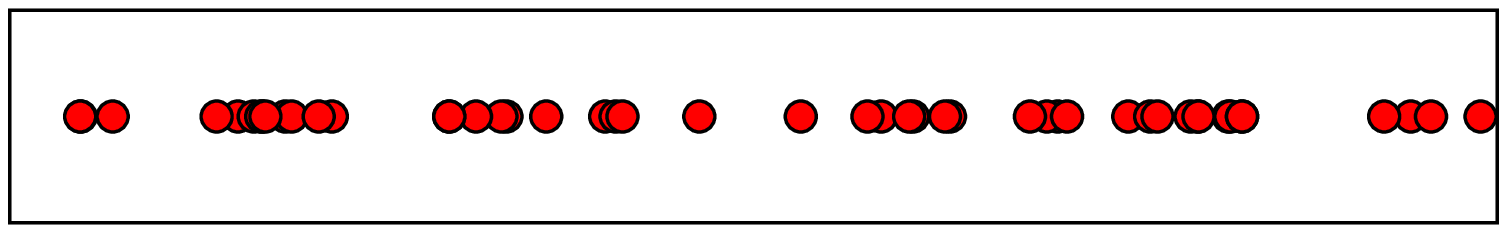}\hspace{0.05\textwidth}
\includegraphics[width=0.45\textwidth]{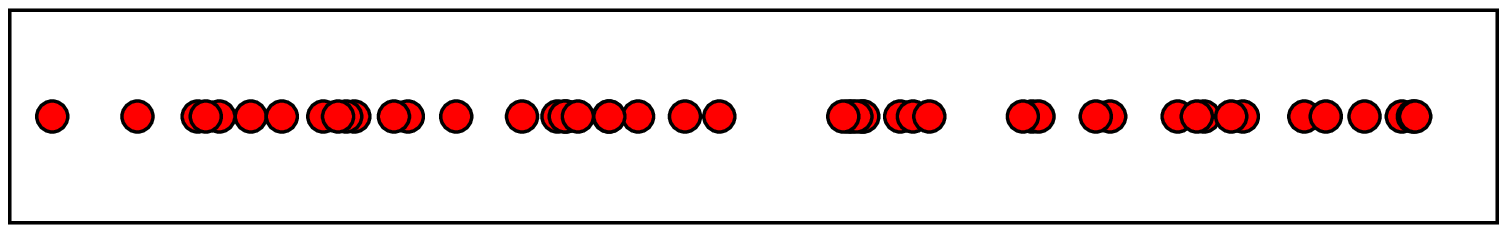}\newline
\includegraphics[width=0.45\textwidth]{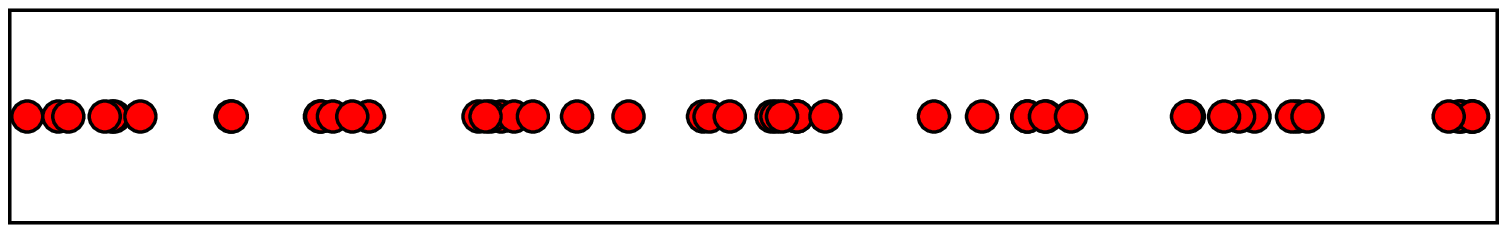}
\caption{(Color online)  Portions of numerically-constructed hyperuniform point 
patterns with $\chi = 0.1$ and small-$k$ exponential scalings (upper left) $\alpha = 0.5$, 
(upper right) $\alpha = 1.0$, and (lower) $\alpha = 2.0$.}\label{config1}
\end{figure}

\begin{figure}[!t]
\centering
\includegraphics[width=0.45\textwidth]{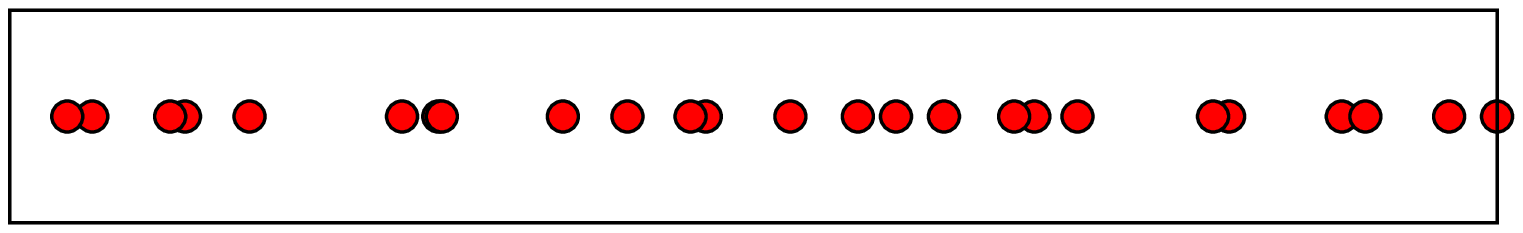}\hspace{0.05\textwidth}
\includegraphics[width=0.45\textwidth]{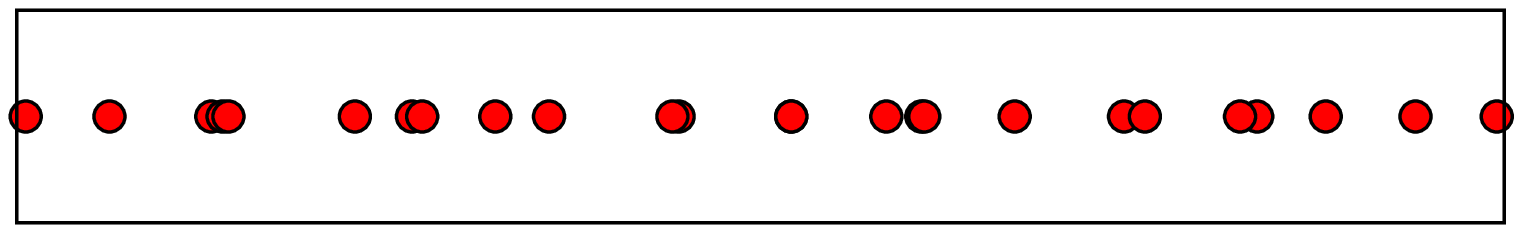}\newline
\includegraphics[width=0.45\textwidth]{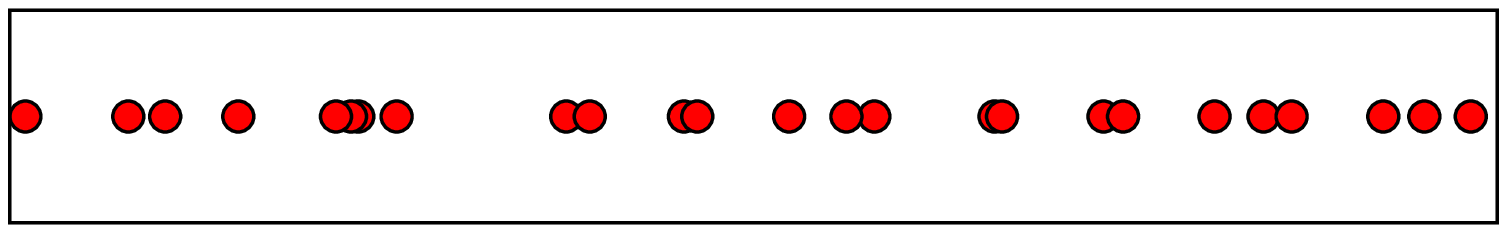}
\caption{(Color online)  Portions of numerically-constructed hyperuniform point 
patterns with $\chi = 0.35$ and small-$k$ exponential scalings (upper left) $\alpha = 0.5$, 
(upper right) $\alpha = 1.0$, and (lower) $\alpha = 2.0$.}\label{config2}
\end{figure}

It is interesting to note the substantial differences in the structure factors of the 
systems for $\chi = 0.1$ and $\chi = 0.35$, particularly for unconstrained 
wavevectors.  For $\chi = 0.1$, the structure factor exhibits an unusually slow 
decay to its asymptotic value of unity; we have fit the large-$k$ region of the 
structure factor with an asymptotic fit of the form $1+\beta/k^{\gamma}$ and 
have found a power-law decay $\gamma = 1$.  This behavior is due to the 
local clustering of particles as expected from the small-$r$ region of the pair 
correlation function in Fig. \ref{chi01g2}.
%[[I NEED TO CHECK THE CONSISTENCY OF THE LARGE-K FIT OF S(K) WITH THE SMALL-R REGION OF G2.]]  
This effect can be directly observed in Figs. \ref{config1} and \ref{config2}, 
which provide illustrative portions of our numerically-constructed hyperuniform 
point patterns at $\chi = 0.1$ and $\chi = 0.35$.  As has been previously 
reported in the literature \cite{UcToSt06, BaStTo08}, increasing the fraction of 
constrained degrees of freedom in the many-particle system has the effect of 
imposing greater local order in the form of an effective short-range repulsive interaction.  
By increasing $\chi$ from $0.1$ to $0.35$, the relative influence of the constrained 
wavevectors on the pair correlation function increases, suppressing the formation 
of local clusters.  However, we also observe that as the exponent $\alpha$ 
controlling the small-wavenumber region of the structure factor decreases 
(equivalently, as anomalous local number density fluctuations appear), this 
effective repulsion between particles becomes noticeable weaker, manifested 
in the pair correlation function by larger values of $g_2(0)$.  This behavior 
suggests that anomalous hyperuniform point patterns possess greater 
variability in their local structures, particularly with regard to the shapes 
and sizes of voids between particles.  

\begin{figure}[!t]
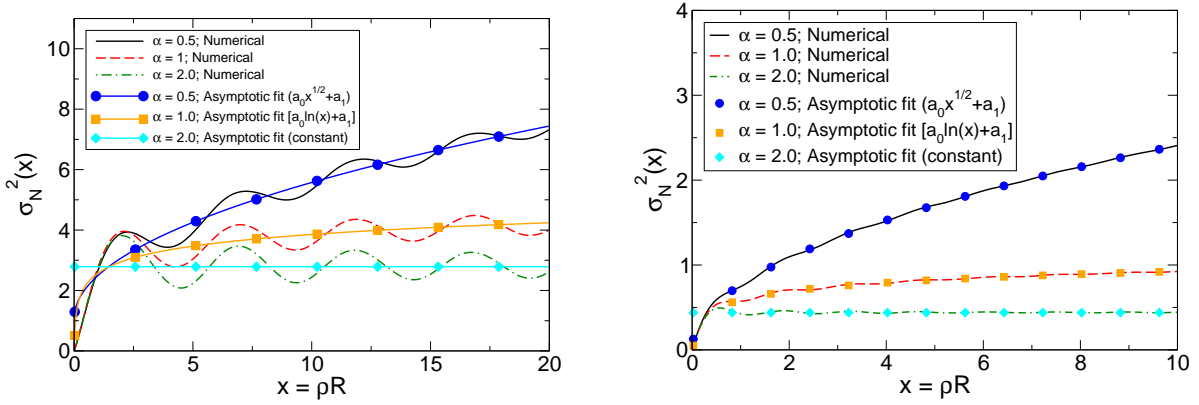

\centering
\includegraphics[width=0.45\textwidth]{Fig9A}\hspace{0.05\textwidth}
\includegraphics[width=0.45\textwidth]{Fig9B}
\caption{(Color online)  Left panel:  Number variances $\sigma^2_N(R)$ for 
numerically-constructed hyperuniform point patterns with $\chi = 0.1$.  Right panel:  
Corresponding number variances for $\chi = 0.35$.}\label{NV}
\end{figure}

We have verified the expected asymptotic behaviors of the number variance for our 
numerically-constructed point patterns as shown in Fig. \ref{NV}.  The asymptotic 
scalings of these fluctuations exactly correspond to their theoretical predictions.  In 
particular, we have provided the first example of a hyperuniform point pattern for 
which the asymptotic number variance grows more slowly than the volume of an 
observation window but faster than a logarithmic scaling.  Interestingly, the local 
clustering of points at $\chi = 0.1$ generates strong oscillations in the number 
variance that persist for several nearest-neighbor distances.  In contrast, these 
local oscillations essentially vanish after two nearest-neighbor distances at $\chi = 0.35$, 
reflecting the strong constraints placed on the local structure by the small-wavenumber 
region of the structure factor.  

%[[EITHER HERE OR IN THE DISCUSSION SECTION, PROVIDE A PHYSICAL EXPLANATION FOR ANOMALOUS FLUCTUATIONS IN TERMS OF SNAPSHOTS OF THE LOCAL STRUCTURE]].

\begin{figure}[!t]
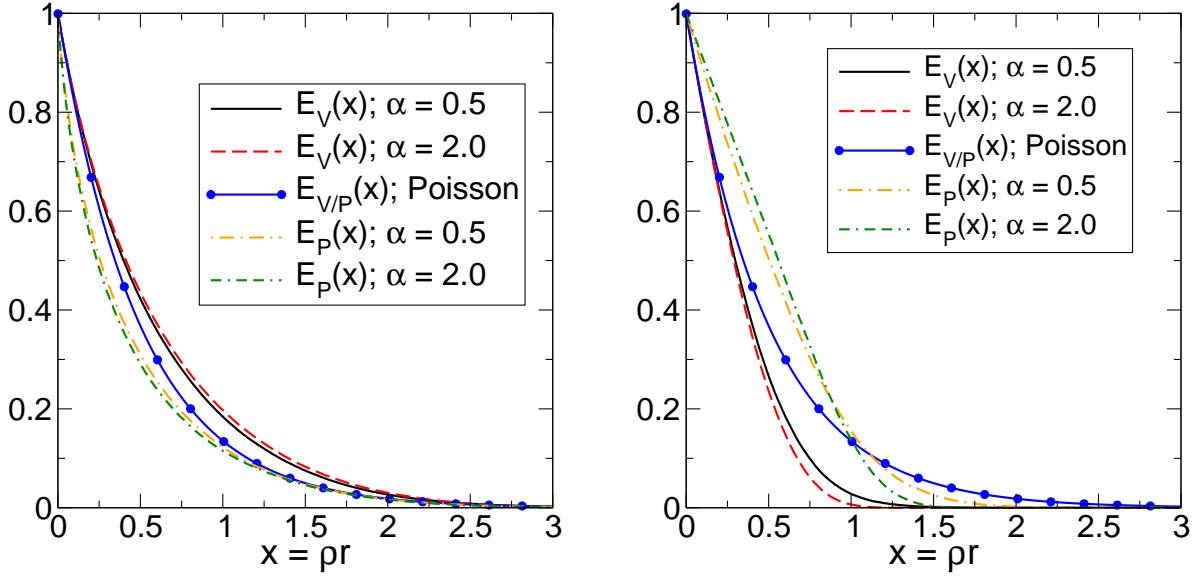

\centering
\includegraphics[width=0.45\textwidth]{Fig10A}\hspace{0.05\textwidth}
\includegraphics[width=0.45\textwidth]{Fig10B}
\caption{(Color online)  Left panel:  Void- and particle-exclusion probabilities for the 
numerically-constructed hyperuniform point patterns ($\chi = 0.1$) along with the 
reference curve for the Poisson point process.  Right panel:  Corresponding 
functions for $\chi = 0.35$.}\label{EvEpfig}
\end{figure}

Our calculations for the void and particle exclusion probabilities of these systems, 
shown in Figure \ref{EvEpfig}, demonstrate previously-unobserved statistical properties 
for hyperuniform point patterns.  For a Poisson point pattern, one has the result that 
\begin{equation}
E_V(r) = E_P(r) = \exp[-\rho v(r)],
\end{equation}
implying that the exclusion correlation function $\eta(r) = 1$ for all $r$.  This result 
follows from the absence of interparticle correlations for the process and the underlying 
Poisson probability 
distribution for the number of particles within an arbitrary compact set.  
Gabrielli and Torquato \cite{GaTo04} have provided strong arguments to suggest 
that for any hyperuniform point pattern, the void exclusion probability should 
asymptotically decay faster than for a Poisson point process.  This behavior implies 
that arbitrarily large 
cavities within the system, while not prohibited by the constraint of hyperuniformity, 
are expected to be significantly rare events owing to the underlying regularity of the 
global structure of the pattern.  
It is therefore not unreasonable to expect the functional form of $E_V(r)$ for the Poisson 
point process to provide an upper bound on the exlcusion probability of any hyperuniform 
point pattern, and this observation is indeed rigorously true for point 
patterns generated from fermionic particle distributions (so-called determinantal point 
processes) \cite{ToScZa08}.  More generally, the Poisson result will place an upper 
bound on $E_V$ for any point pattern with $n$-particle correlation 
functions $g_n \leq 1$ for all $n$.   

For $\chi = 0.1$, we observe the unusual property that $E_V(r)$ is greater than the 
Poisson result for all values of $r$ that can be reliably determined from numerical 
simulation.  It is instead the \emph{particle} exclusion probability function 
that is bounded from above by the Poisson curve.  To understand this discrepancy, 
we first note that $E_V(r)$ and $E_P(r)$ are rigorously bounded from below by 
[c.f. \eqref{lbound}] \cite{ToScZa08}
\begin{align}
E_V(r) &\geq 1-\rho v(r)\label{Evlower}\\
E_P(r) &\geq 1-Z(r)\label{Eplower},
\end{align}
where $Z(r)$ is the \emph{cumulative coordination number}
\begin{equation}
Z(R) = \rho \int_{\mathbb{R}^d} \Theta(R-r) g_2(\mathbf{r}) d\mathbf{r}.
\end{equation}
These bounds become sharp at low density or small $r$.  Therefore, while 
$E_V(r)$ is related to the geometry of a cavity within the void space, the 
particle exclusion probability $E_P$ depends explicitly on the local coordination structure
of the underlying point process.  
%Since hyperuniformity imposes local uniformity amongst the points compared to a Poisson point process, the cumulative coordination number at small $r$ will be less than the corresponding value for the Poisson point pattern,  
%reflecting the local effective repulsion amongst points within the system.  
%The above statement is INCORRECT because hyperuniformity does not imply the presence of local regularity, only staticity with respect to local observation windows.

To elucidate further the relationship between the local coordination structure 
and the void statistics, we can consider a modification of the number variance problem, 
whereby one measures fluctuations in the number of points within an observation 
window \emph{centered on a point of the point process}.  
Let $N_P^{(i)}(R)$ denote this quantity; it can be represented as
\begin{equation}
N_P^{(i)}(R) = \sideset{}{^\prime}\sum_{j=1} \Theta(R-\lVert\mathbf{r}_j-\mathbf{r}_i\rVert),
\end{equation}
where the prime on the summation means that particle $i$ is excluded.  
The average value of this random variable is
\begin{equation}
\langle N_P^{(i)}(R)\rangle = \rho\int_{\mathbb{R}^d} g_2(\mathbf{r}) 
\Theta(R-\lVert\mathbf{r}\rVert) d\mathbf{r} = Z(R).
\end{equation}   
The cumulative coordination number therefore measures the local number 
density within \emph{the particle space}.  It follows that when 
$Z(R) \leq \rho v(R)$,  $\langle N_P^{(i)}(R)\rangle$, the average number of points 
in the particle space
within an observation window of radius $R$, is less than or equal to $\langle N_V(R)\rangle$, 
the average number of points of the process within a window in the void space.  
This behavior then implies that 
the points are more greatly dispersed within the particle space, and $E_P(R) \geq E_V(R)$; 
equivalently, $\eta(R) \geq 1$.  Note that this analysis is consistent with the lower 
bounds \eqref{Evlower} and \eqref{Eplower} on the exclusion probability
functions.  

Conversely, for the case where $Z(R) \geq \rho v(R)$, we have that 
$\langle N_P^{(i)}(R)\rangle \geq \langle N_V(R)\rangle$, which suggests that 
the points are more closely located within the particle space, leaving larger 
cavities within the void space.  
We therefore conclude that $E_P(R) \leq E_V(R)$ [$\eta(R) \leq 1$], and the 
point process should exhibit local clustering among points.  These claims are 
also consistent with our results for the exclusion probability functions in Fig. \ref{EvEpfig},
whereby we observe a transition from $\eta(R) < 1$ at $\chi = 0.1$ to 
$\eta(r) > 1$ for $\chi = 0.35$.  Indeed, the configurations at $\chi = 0.1$ 
exhibit substantial clustering among points (c.f. Fig. \ref{config1}). 

Our arguments can be extended by examining the scaling of the 
configurational degeneracy with the fraction of constrained degrees of freedom 
$\chi$.  
Here we measure this degeneracy by calculating the entropy 
(logarithm of the degeneracy) of the system relative to an ideal gas 
of $N$ particles in a volume $V$ on the line.  
For the ideal gas, we coarse-grain the system by dividing the volume 
$V$ into $M \gg 1$ cells such that no more than one particle occupies 
each cell with probability one, thereby 
representing the degeneracy associated with the $dN$ translational 
degrees of freedom as a combinatorial occupancy problem.  
The size of a cell determines the length scale, meaning without loss of 
generality that we need only consider the regime $\rho = N/M \ll 1$.  
Assuming that the particles are indistinguishable, the number of 
configurations $\Omega$ available to the system is
\begin{equation}
\Omega = \frac{M!}{(M-N)! N!}.
\end{equation}
Since the underlying distribution of particles is uniform within the cells, 
Boltzmann's formula for the entropy gives (with $k_B = 1$)
\begin{equation}
S = \ln \Omega = \ln\left[\frac{M!}{(M-N)! N!}\right],
\end{equation}
which for large $M$ and $N$ becomes  
\begin{align}
S &= M\ln M - M - (M-N)\ln(M-N) + (M-N) - N\ln N + N\\
&= -M\ln(1-N/M) + N\ln(M/N) + N\ln(1-N/M).
\end{align}
Under the assumption that $N/M \ll 1$, we have the following result for the 
entropy per particle $\overline{S}_{\text{ideal}} = S_{\text{ideal}}/N$:
\begin{equation}
\overline{S}_{\text{ideal}} = 1-N/M + \ln(M/N) \approx 1-\ln(\rho).
\end{equation}
The entropy of the ideal gas is therefore large and positive as expected.  
For the density regime $\rho \ll 1$ that we have in mind, one can simplify 
further by taking $\overline{S} \approx -\ln(\rho)$, which diverges to $+\infty$ for small $\rho$.    

Suppose now that we constrain $K$ degrees of freedom, where $K \ll N$.  
This construction correponds to making a small perturbation away from the 
ideal gas configuration with $N-K$ degrees of freedom still available to the 
many-particle system.  
We again divide the volume $V$ into $M$ cells of unit length ($N/M \ll 1$).  
For sufficiently small values of $\chi$ (i.e., near the ideal gas), we may assume 
that the length scale of the 
effective repulsion between particles is negligible compared to the 
cell size, meaning that the $N-K$ unconstrained particles may be distributed freely 
among the $M$ cells.  Note that the $K$ constrained degrees of freedom are explicitly 
determined once these particles have been placed.  
The number of microstates available to the system is then
\begin{equation}
\Omega = \frac{M!}{(M-N+K)!(N-K)!},
\end{equation}
and the configurational entropy is
\begin{equation}
S = \ln\Omega = M\ln\left(\frac{M}{M-N+K}\right) + K \ln\left(\frac{N-K}{M-N+K}\right) 
+ N\ln\left(\frac{M-N+K}{N-K}\right).
\end{equation}  
Defining the fraction of constrained degrees of freedom $\chi = K/N$, we may write 
for the entropy per particle $\overline{S} = S/N$:
\begin{equation}
\overline{S} = (1-M/N-\chi)\ln(1-N/M+\chi N/M) + (1-\chi)\ln(M/N)-(1-\chi)\ln(1-\chi)
\end{equation}
where $\chi \ll 1$.
For $N/M \ll 1$, this expression simplifies as
\begin{equation}
\overline{S} = 1-\chi - (1-\chi)\ln(1-\chi) + (1-\chi)\ln(M/N) - (N/M)(1-\chi)^2.
\end{equation}
The last term in this expression is negligible within the density regime where 
$\overline{S}_{\text{ideal}} \approx \ln(M/N) = \ln(1/\rho)$.  
Since $\overline{S}_{\text{ideal}}$ is large and positive compared to the first 
terms of this result, we have the expected scaling
\begin{equation}
\frac{\overline{S}}{\overline{S}_{\text{ideal}}} \approx 1-\chi,
\end{equation}
suggesting a roughly \emph{linear} decrease in the entropy of the system 
for small values of $\chi$.  

By increasing the parameter $\alpha$, we expect also to increase the 
rate at which $\overline{S}\rightarrow 0$ with respect to $\chi$ since 
higher values of $\alpha$ are associated with larger effective radii around 
the constrained particles.  However, since we require that our configurations 
are hyperuniform, there are additional implicit constraints on the 
``unconstrained'' degrees of freedom.
Formally, hyperuniformity requires that the local structure of a point 
process approach the global structure over sufficiently short length scales, 
on the order of several nearest-neighbor distances.  This behavior is typically associated with 
a highly regularized distribution of points, such as with a Bravais lattice.  
However, our results for $E_V$ at $\chi = 0.1$ suggest an alternative 
mechanism by which hyperuniformity can be achieved in a point pattern.  
Specifically, the fact that $\eta(r) < 1$ 
is consistent with local clusters of particles that are \emph{globally} 
regularized by an increased probability of finding sufficiently large voids to separate them.  
In this case, the appearance of these large voids external to the 
particle space is apparently essential to enforce hyperuniformity of the point 
pattern by overcoming the highly inhomogeneous local structure of the clusters.  
In the context of our analysis above, for small perturbations from the ideal gas, the 
high configurational degeneracy that remains after constraining only a few 
degrees of freedom implies that highly disordered configurations are most likely 
to appear from our numerical constructions.  However, with the added implicit 
constraint of hyperuniformity
the system will sacrifice local structural regularity for clustering of points that 
are globally separated by sufficiently large voids, resulting in a negatively 
correlated exclusion correlation function.  

\begin{figure}[!tp]
\centering
\includegraphics[width=0.5\textwidth]{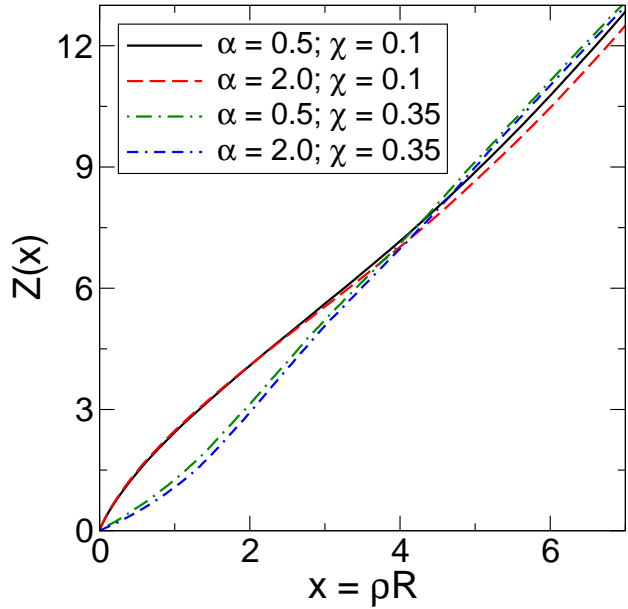}
\caption{Cumulative coordination numbers $Z(R)$ for numerically-constructed 
hyperuniform point patterns with small-wavenumber exponents $\alpha = 0.5$ 
and $\alpha = 2.0$.  The fractions of constrained degrees of freedom 
are $\chi = 0.1$ and $\chi = 0.35$.}\label{Z01035}
\end{figure}
Figure \ref{Z01035} highlights this behavior 
by examining the cumulative coordination numbers $Z(R)$ at $\chi = 0.1$ and 
$\chi = 0.35$ for our systems.   Although at $\chi = 0.1$ the $\alpha = 2$ system 
exhibits greater clustering at small-$r$ than for $\alpha = 0.5$, at large $r$ 
this trend reverses, which is consistent with the appearance of larger voids with 
higher probability and supports our claim that these voids serve to regularize the 
global structures of the systems.  Using the lower bound \eqref{Eplower} on the particle
exclusion probability function, we also observe explicitly the effect of the locally 
clustered structure on the small-$r$ particle exclusion probability.   

Upon reaching $\chi = 0.35$, we recover the usual behavior associated with hyperuniformity.  
By constraining a sufficient number of degrees of freedom, the effective interparticle repulsions 
induced by the collective coordinate constraints
control the small-$r$ region of the pair correlation function (and, therefore, the cumulative 
coordination number), prohibiting local cluster formation.  The regularizing factor in this 
case is therefore the distribution of voids within the particle space,
contained in $E_P(r)$.  Indeed, the void exclusion probability $E_V(r)$ is highly constrained 
by this effective repulsion and decays to zero faster than $E_P(r)$, resulting in a positively 
correlated exclusion correlation function.  By 
increasing the exponent $\alpha$ governing the small-wavenumber region of the structure 
factor, we observe increased regularity in the local structure, corresponding to a decreased 
$Z(R)$ and $E_V(r)$ and a particle exclusion correlation function
$E_P(r)$ that decays more rapidly to zero, in perfect accordance with the aforementioned 
void-space criterion on hyperuniformity put forth by Gabrielli and Torquato \cite{GaTo04}.  
It is also noteworthy that increased void-space constraints associated 
with increased $\alpha$ are consistent with the behavior of the structure factor for MRJ 
hard-sphere packings, about which we have more to say in Section VI.

\section{Concluding remarks and discussion}

We have provided the first known constructions of disordered hyperuniform many-particle 
ground states possessing anomalous local density fluctuations.  Such systems are defined 
by a number variance $\sigma^2_N(R)$ that asymptotically scales faster than 
the surface area of an observation window but slower than the window volume.  
By controlling the collective density variables associated with the underlying point pattern, 
we have also be able to 
probe the relationship between interparticle correlations and constraints on the local 
coordination structure.  Specifically, we have provided detailed statistics to measure 
the distribution of the \emph{void space} external to the particles,
including measurements of the void and particle exclusion probabilities.  

Under sufficiently low constraints on the system, our numerically constructed 
many-particle distributions exhibit substantial clustering, resulting in a highly 
inhomogeneous
local structure.  However, on the global scale of the system as measured by 
asymptotic local density fluctuations, these local clusters are separated by 
comparatively large interparticle voids, thereby regularizing the
microstructure and preserving the constraint of hyperuniformity that we impose.  
Indeed, this effect becomes more pronounced upon passing from the ``anomalous'' 
regime of hyperuniformity to the more usual case, where $\sigma^2_N(R) \sim R^{d-1}$
asymptotically (i.e., for all periodic point patterns, quasicrystals with Bragg peaks, 
and disordered systems with pair correlations decaying exponentially fast) \cite{ToSt03, ZaTo09}.  
Upon increasing the fraction of constrained degrees of freedom within the system, 
we are able to preclude this clustering affect by reinforcing the effective interparticle 
repulsion imposed by 
our targeted structure factor $S(k)$.  Furthermore, we have shown that this effective 
repulsion becomes increasingly more pronounced as the exponent $\alpha$ 
governing the small-wavenumber scaling of $S(k)$ is increased.  

It follows from these observations that one can formally define an effective repulsive 
radius around each point within a hyperuniform point pattern.  However, so long as 
this radius is not substantially large compared to the expected interparticle spacing
$\rho^{-1/d}$, i.e., in the absence of microstructural constraints, clustering effects can 
still dominate the interparticle correlations and the local coordination structure.  
We have shown that this effect is entropically favorable since slight deviations
from the ideal gas are still associated with an exponentially large configurational degeneracy.  
However, this degeneracy is expected to increase rapidly upon constraining a sufficient 
number of degrees of freedom, or, equivalently, increasing the
effective repulsive radius.  We have shown that this loss of configurational degeneracy is 
associated with a highly-constrained void space distribution, which can be considered as 
a signature of predominantly 
``repulsive'' hyperuniform point patterns.  

Our results have particular implications for understanding the appearance and nature of 
hyperuniformity in MRJ packings of hard spheres.  
It is interesting to note that equilibrium distributions of hard spheres are known not to be 
hyperuniform \cite{ToSt03} except
at the close-packed density, at which point the system freezes into a crystal with long-range 
order.  MRJ hard-sphere packings are therefore unique in that they are 
\emph{nonequilibrium} systems that are uniformly mechanically rigid, and it is 
this rigidity that has been shown to be essential for the onset of hyperuniformity 
(with logarithmic asymptotic local density fluctuations) \cite{ZaJiTo10}.  Importantly, 
rigidity places severe geometric constraints on the local arrangements of particles
\cite{DoToSt05} and has been shown to regularize the void space distribution on the 
global scale of the microstructure \cite{ZaJiTo10}. We have demonstrated in this work 
that by decreasing the exponential form of the structure factor 
within the small-wavenumber region,
these void-space constraints are relaxed in accordance with the decreased effective 
radius surrounding the particles.  Since MRJ packings are \emph{maximally} disordered 
among all strictly jammed packings, it follows that such systems
must already possess the maximal number of degrees of freedom consistent with the 
geometric constraints of strict jamming.  Therefore, \emph{any} increase in the distribution 
of the void sizes is inconsistent with these same constraints, 
highlighting why exponents $\alpha < 1$ in the small-wavenumber scaling of $S(k)$ 
have never been observed for such systems.  

Our work has also raised a number of interesting questions related to the physics of 
collective coordinate constraints.  The mathematical properties associated with 
collective coordinates are surprisingly subtle and have only partially 
been explored in the literature \cite{FaPeStSt91, UcStTo04}.  Constraining a 
collective density variable, including, for example, either a complete suppression 
to zero or fixing its magnitude, results in a highly nonlinear
equation relating the components of the particle positions $\{\mathbf{r}_j\}$.  
For higher-dimensional systems, it has been previously observed \cite{BaStTo09} 
that these nonlinear equations have a tendency to ``couple'' in such a way that
one must go beyond $\chi = 0.5$ to crystallize the system, meaning that one cannot 
simply count constraints and degrees of freedom.  It is an open problem to determine 
analytically the relationship between $\chi$ 
and the ``true'' number of constrained degrees of freedom; recent work has examined 
the fraction of normal modes with vanishing frequency as a more appropriate indicator 
of the latter \cite{BaStTo09}.
We have provided some analysis here in one dimension to suggest how the configuration 
space is constrained with increasing $\chi$ by determining the entropy 
for small deviations from the ideal gas.  Certainly for higher values of $\chi$ our simple 
linear scaling will break down; characterizing the deviations from this linear behavior, 
especially in higher dimensions, is an attractive problem 
warranting further consideration.


\begin{thebibliography}{}
\bibitem{FN1}
Equation \eqref{seventeen} should rigorously be interpreted in the thermodynamic 
limit since the $\mathbf{k} = \mathbf{0}$
mode contains an $\mathcal{O}(1/N)$ correction to the long-range behavior of $g_2$.
\bibitem{FN3}
For the one-dimensional case where $S_0(k) = 0$ for all $k \in \mathcal{Q}$, $\chi = 1/2$ as 
defined in \eqref{chidef}
corresponds to a loss of all degrees of freedom.  To see this, note that $S_0(k) = 0$ is 
equivalent to requiring $\hat{\rho}(k) = 0$, which places two independent constraints, 
corresponding 
to real and imaginary parts, on the $N$ degrees of freedom \cite{FaPeStSt91}.
\bibitem{FNhighd}
For reasons that we discuss in Section VI, in higher dimensions one cannot simply use 
the number of vanishing 
collective coordinates to predict the value of $\chi$ at which the disorder-order phase 
transition occurs.  
Numerical work in the literature suggests that the threshold values of $\chi$ in two and 
three dimensions are,
respectively, 0.77990 and 0.50066 \cite{UcToSt06, BaStTo08}.
\bibitem{FN4}
The case $S_0(\mathbf{k}) = 0$ can be recovered either as the limit $\alpha \rightarrow 
+\infty$ or by taking $D = 0$.
\end{thebibliography}
\end{document}